\documentclass[twocolumn,aps,prd,10pt]{revtex4-1}

\usepackage{amsmath}
\usepackage{amssymb}
\usepackage{amsfonts}
\usepackage{mathrsfs}

\newcommand{\g}{\mathrm{g}}

\newcommand{\LLL}{\mathscr{L}}
\newcommand{\HHH}{\mathscr{H}}
\newcommand{\E}{\text{E}}
\newcommand{\G}{\text{G}}
\renewcommand{\S}{\text{S}}

\newcommand{\M}{\text{M}}
\newcommand{\diag}{\text{diag}}

\newcommand{\ovr}[1]{\overline{#1}}
\newcommand{\und}[1]{\underline{#1}}
\newcommand{\DDD}{\text{D}}
\newcommand{\A}{\mathcal{A}}
\newcommand{\FF}{\mathcal{F}}
\newcommand{\HH}{\mathcal{H}}

\newcommand{\CC}{\mathcal{C}}
\newcommand{\DD}{\mathcal{D}}
\newcommand{\GG}{\mathcal{G}}
\newcommand{\dd}{\text{d}}

\newcommand{\ppp}{\\[3pt]}
\newcommand{\nppp}{\nonumber\\[3pt]}
\newcommand{\bppp}{\\[-9pt]}

\begin{document}

\title{Closing the gaps in quantum space and time: Conformally augmented gauge\\ structure of gravitation}

\author{Charles H.-T. Wang}
\email{c.wang@abdn.ac.uk}

\author{Daniel P. F. Rodrigues}

\affiliation{\vspace{3pt} \mbox{Department of Physics, University of Aberdeen, King's College, Aberdeen AB24 3UE, United Kingdom}\vspace{5pt}}

\begin{abstract}
A new framework of loop quantization that assimilates conformal and scale invariance is constructed and is found to be applicable to a large class of physically important theories of gravity and gravity-matter systems. They include general relativity and scale-invariant scalar-tensor and dilaton theories. Consequently, matter to be coupled to such theories is restricted to be conformal or scale invariant. Standard model-type systems naturally fall into this category. The new loop quantization follows from a novel conformally generalized Holst action principle. In contrast to standard loop quantum gravity, the resulting quantum geometry is not beset by the Immirzi ambiguity and has no definitive area gaps within the considered large class of theories of gravitation. As an additional feature, the scale invariance gives rise to a conserved Weyl current and we discuss briefly its possible implication on the problem of time in quantum gravity.
\vspace{15pt}
\\
\href{https://doi.org/10.1103/PhysRevD.98.124041}
{DOI: 10.1103/PhysRevD.98.124041}
\hfill
\href{https://arxiv.org/abs/1810.01232}
{arXiv:1810.01232}
\end{abstract}

\maketitle

\section{Introduction}
\label{sec:a}

One of the most ubiquitous questions in physics is the scale of a given system and arguably the most fundamental of such issues is the basic size of the building blocks of space itself. Loop quantum gravity (LQG) \cite{Thiemann2008, Rovelli2011, Ashtekar2017} is a promising candidate to unify quantum theory and general relativity.
En route to this final goal, apart from certain outstanding technical issues, there are intriguingly fundamental challenges to LQG, which we address in this work: (i) the persistent Immirzi ambiguity \cite{Immirzi1997a} responsible for the uncertain microscopic units of quantized geometric quantities such as areas must be satisfactorily resolved \cite{Rovelli2011, Achour2017}; (ii) the quantized geometry should at least support and at best imply the physical forms of matter, particularly in the standard model (SM) \cite{Thiemann1998, Meissner2007, Bars2014, Campiglia2017}; and (iii) the consequent cosmic evolution should be compatible with or indeed account for the large scale structures and scale independent fluctuations of the observed Universe \cite{Ferreira2016, Maeder2017a, Maeder2017b, Midy1999, Midy2003}.

Ostensibly, these three points have completely different guises. But they share a similar geometric attribute in common, with a strong hint of a universal scaling symmetry. While the scale invariance in cosmology and the near conformal invariance of the SM have been extensively studied \cite{Ferreira2017, Ferreira2018}, there is a lack of general consensus on its analogous role in quantum gravity.
The present formalism of LQG does not have a scale invariance. To the contrary, it predicts the existence of small but finite units of geometric measures like areas and their quantum gaps \cite{Rovelli1995}. Because of the Immirzi ambiguity, it remains uncertain what the related scale should be. While not affecting normal macroscopic measures, this is potentially serious since, if true, loop quantum cosmology would imply the big bang had a passage to a hypothetical previous universe bearing further observational ramifications \cite{Bojowald2005, Agullo2016}.
Furthermore, the effect of fundamental scales on the early Universe may have profound implications on quantum gravity phenomenology such as gravitational decoherence and stochastic gravitational waves \cite{Oniga2016a, Oniga2017b, Quinones2017} in the late Universe.

Here we present a novel generalization of LQG using a conformal gauge structure. Through a new conformally augmented analysis, we derive the conformally covariant form of the extrinsic curvature that enables us to construct the conformally extended Ashtekar-Barbero variables \cite{Ash86, Bar95}.
A conformal Immirzi parameter naturally arises in this construction with a
value related to the choice of conformal frames.
A key feature is the existence of the conformal constraint acting as the canonical generator of conformal transformations that shift the value of the conformal Immirzi parameter.
The presented theoretical framework is amenable to loop quantization. In particular, the implementation of the conformal constraint using schemes developed in Refs.~\cite{Thiemann1996, Lewandowski2016} may lead to the unitary equivalence of loop representations with different values of the conformal Immirzi parameter.

We start our analysis in the coordinate-independent manner using the lagrange of modern differential forms and exterior algebra. When passing over to the canonical analysis, a $(3+1)$-decomposition of spacetime coordinates is introduced in a time gauge.
The main focus of this work is a new conformally augmented dynamical structure which is amenable to generic loop quantization techniques. Accordingly, our discussion on quantization will be restricted to key steps. Nonetheless, we show how our new approach may significantly alters implications from the modified LQG.

We use fairly standard geometric notation summarized here for clarity.
In units where $\kappa$ $(=8\pi G/c^4) = c = 1$,
we consider the tetrad 1-forms
${e}^I = e_\alpha{}^I \dd x^\alpha$  with
$I=0,1,2,3$
and its dual tangent vectors
$\tilde{e}_I = e^\alpha{}_I \partial_\alpha$
with contraction
${e}^I(\tilde{e}_J)=\delta^I_J$
and
$\epsilon=\det(e_\alpha{}^I)$.
Here spacetime coordinates are
$(x^\alpha)=(t, x^a)$ with
$\alpha=0,1,2,3$ and
$a=1,2,3$.
The Levi-Civita (LC) antisymmetric symbols are denoted by
$\epsilon_{IJKL}$
with $\epsilon_{ijk} = \epsilon_{0ijk}$ for $i,j,k=1,2,3$
and
$\epsilon_{0123}=\epsilon_{123}=1$.
The spacetime metric tensor is
$\g = \eta_{IJ}\, e^I \otimes e^J$
using the Minkowski metric
$\eta_{IJ}=\diag(-1,1,1,1)$
with
$g_{\mu\nu} = e_\mu{}^I e_\nu{}_I$
and
$g = \det(g_{\mu\nu})=-\epsilon^2$.
The metric compatible connection 1-forms $A_{IJ}$ independent of the tetrad are used to define the covariant exterior derivative $\DDD$. In particular, the torsion 2-forms of $A_{IJ}$ are given by
$
T^{I}
=
\DDD\, e^I
=
\dd e^I + A^I{}_J \wedge e^J
$
using the exterior product $\wedge$ and derivative $\dd$,
satisfying the first Bianchi identity
$
\DDD\, T^{I}
=
F^I{}_J \wedge e^J
$
where
$
F_{IJ}
=
\dd A_{IJ} + A_{IK} \wedge A^K{}_J
$
are the curvature 2-forms of $A_{IJ}$  \cite{Tucker1987}. The coordinate components of $A_{IJ}$ and $F_{IJ}$ are denoted by
$A_{\alpha IJ}$ and $F_{\alpha\beta IJ}$ and similarly for other quantities.

\section{Conformal Holst action for general relativity}
\label{sec:b}

Let us begin with the simultaneous conformal gauge changes of tetrad and connection 1-forms
\begin{eqnarray}
\ovr{e}^I
&=&
\phi\, {e}^I
,\quad
\ovr{A}_{IJ}
=
A_{IJ} + \phi_{IJ}
\label{be}
\end{eqnarray}
for any positive scalar field $\phi$ regarded as the primary conformal factor. Likewise, any conformally transformed quantity induced by the above will be denoted with an overline.
Here the 1-forms
$
\phi_{IJ}
=
-2\,\dd(\ln\phi)(\tilde{e}_{[I})\,  e_{J]}
$
are uniquely determined to guarantee the conformal gauge covariance
$\ovr{\DDD\, e^I}=\phi\,\DDD\, e^I$.

To explore the effect of relative conformal changes, i.e. the ``relativity of conformal frames'', we will also consider similar conformal changes of any quantity with respect to another (secondary) conformal scalar field $\theta$, denoted by an underline, with analogous $\theta_{IJ}$. This enables us to construct a conformally extended Holst action $S[A,e,\phi,\theta]$ as
\begin{eqnarray}
\hspace{-10pt}
S
&=&
\frac1{2}\int
\Big[
\ovr{F_{IJ} \wedge \star (e^I\wedge e^J)}
-
\und{F_{IJ} \wedge e^I \wedge e^J}\,
\Big]
\label{S}
\end{eqnarray}
consisting of the first Palatini and second Holst terms \cite{Holst1996}, nontrivially transformed with the primary and secondary conformal factors $\phi$ and $\theta$ respectively.

The action \eqref{S} has recently been conjectured to yield conformally extended Ashtekar-Barbero connection variables free of the longstanding Immirzi ambiguity \cite{Veraguth2017}.
To proceed, we first derive the covariant field equations by varying the action \eqref{S} with compact support, which turn out to be equivalent to Einstein's equations for general relativity (GR). We will then demonstrate that such conformal Ashtekar-Barbero variables indeed arise from the conformally extended canonical analysis.

Denoting the variation with respect to  $\sqcup$ by $\delta_{\sqcup}$, we find the variations of the two terms in Eq.~\eqref{S} with respect to the connection $A_{IJ}$ yields
\begin{eqnarray*}
\delta_A S
&=&
\frac12\,
\delta {A}_{IJ}
\wedge
\big[
\phi^{2} \DDD\star (e^I\wedge e^J)
-
\theta^{2} \DDD(e^I\wedge e^J)
\big]
\end{eqnarray*}
where $\star$ is the Hodge dual.
Consequently, the connection is torsion-free, given by the LC connection $\Gamma_{IJ}$, i.e.
\begin{eqnarray}
\DDD\,e^I
&=&
0
\,
\iff
\,
A_{IJ}
=
\Gamma_{IJ} .
\label{A2Gam}
\end{eqnarray}
Next, the $\phi$-variation yields
\begin{eqnarray*}
\delta_\phi S
&=&
\frac{\delta\phi}{\phi}\,
\ovr{F_{IJ}\!\wedge\!\star (e^I\!\wedge e^J)}
+
\frac{\delta_\phi {\phi}_{IJ}}{2}\!
\wedge
\ovr{\DDD\!\star (e^I\!\wedge e^J)} .
\end{eqnarray*}
Thus, by using Eq.~\eqref{A2Gam}, the second  term of the above vanishes and so the first term must vanish as part of the field equations,
which is equivalent to the vanishing of the scalar curvature
$\ovr{R}$.
This relation is also fulfilled if $\ovr{g}_{\mu\nu}$ satisfies the vacuum Einstein equation, as will be established below.
The $\theta$-variation follows as
\begin{eqnarray*}
\delta_\theta S
&=&
-
\frac{\delta\theta}{\theta}\,\und{F_{IJ}\!\wedge (e^I\!\wedge e^J)}
-
\frac{\delta_{\theta\,} {\theta}_{IJ}}{2}\!
\wedge
\und{\DDD (e^I\!\wedge e^J)} .
\end{eqnarray*}
Again, the second term of the above vanishes by using Eq.~\eqref{A2Gam} and so the first term must vanish as part of the field equations, i.e.
$
{F_{IJ} \wedge (e^I\wedge e^J)}
=0
$
which is also satisfied as a result from the first Bianchi identity with zero torsion.

Last but not least, the variation of the action \eqref{S} with respect to the tetrad yields
\begin{eqnarray*}
\delta_e S
&=&
\frac{\delta e^I}{2}\! \wedge
\big[
\phi\; \ovr{G}_I
-
2\,\theta\,\und{e^J \wedge F}_{IJ}
\big]
\end{eqnarray*}
where the second term also vanishes on account of the first Bianchi identity with zero torsion whereas in the first term $\ovr{G}_I$ are the Einstein 3-forms \cite{Tucker1987}, carrying the same information as the Einstein tensor. This yields the main resulting variational field equation as the Einstein equation for the physical metric $\ovr{g}_{\mu\nu}$, supplemented with the LC spacetime connection and the arbitrary Lagrangian multiplierlike conformal scalar fields $\phi$ and $\theta$.

\section{Canonical analysis of the conformal Holst action}
\label{sec:c}

To derive the promised conformal Ashtekar-Barbero connection variables, we now carry out the canonical analysis of the action \eqref{S}. (Details can be consulted in Supplementary Material Sec.~I \cite{SuppM})
For this purpose, as with the canonical analysis of the standard Holst action \cite{Holst1996}, we adopt a $(3+1)$ coordinate decomposition known as the ``time gauge,'' in which the tetrad is related to the lapse function $N$, shift vector $N^a$, and  triad $e_a{}^i$ by
$
e_\alpha{}^0
=
N\, \delta_\alpha^t
$
and
$
e_t{}^i
=
N^a e_a{}^i
$
with
$e = \det(e_a{}^i) = \epsilon/N$.
As a result, $\tilde{e}_0$ is a unit vector normal to the equal-$t$ hypersurface $\Sigma$ and unit vectors $\tilde{e}_i$ are perpendicular to $\tilde{e}_0$, spanning the tangent space to the spatial hypersurface with metric
$
h_{ab}
=e_a{}^i e_b{}^i
$
and
$ h =\det(h_{ab}) = e^2$.
The densitized triad is then
$E^a_i = e\,  e^a{}_i$
with
$E = \det(E^a_i) = h$.



A close analogy with the Immirzi parameter in standard LQG is obtained, without the loss of generality, by performing a conformal transformation so that of the secondary conformal factor is fixed to be $\theta=1/\sqrt{\beta}$ for some positive constant $\beta$.
However, as will become clear below, its value can still be related to the choice of conformal frames with respect to the primary conformal factor $\phi$, and so we will refer to $\beta$ as the {\it conformal Immirzi parameter}.
Then, the Lagrangian density of the action \eqref{S} becomes
\begin{eqnarray}
\LLL
&=&
\pi_\phi\,\dot{\phi}
+
{\beta}^{-1}{E^a_i}\dot{A}_a^i
+
E^a_i \GG_a^i
\nppp&&
-
N^a E^b_i\,
\HH_{ab}{}_{i 0}
+
\frac{N}{2\sqrt{E}}E^a_i  E^b_j\,
\HH_{ab}{}_{i j}
\label{Lha}
\end{eqnarray}
where
\begin{eqnarray*}
&&
\pi_\phi
\,=
-2\phi\,E^a_i \ovr{A}_{a i 0},
\ppp
&&
A_a^i
=
\beta\phi^2 \ovr{A}_{a i 0}
-
\frac{1}{2}\,
\epsilon_{ijk}
A_{a jk},
\ppp
&&
\GG_a^i
\;=
\frac{\epsilon_{ijk}}{2\beta}
\big(
A_{t jk, a}
+
2A_{t j m} A_{a k m}
-
2A_{t j 0} A_{a k 0}
\big)
\ppp&&
\hspace{40pt}
+
\phi^2
\big(
2\ovr{A}_{[t}{}^{i m} \ovr{A}_{a]\, m 0}
-
\ovr{A}_{t i\, 0, a}
\big),
\ppp
&&
\HH_{ab\,IJ}
=
\phi^2 \ovr{F}_{ab\,IJ}
+
\frac{1}{2\beta}\,
\epsilon_{IJ}{}^{KL}F_{ab\,KL},
\end{eqnarray*}
%
with $\dot{\sqcup}=\partial_t \sqcup$ and $\sqcup_{,a}=\partial_a \sqcup$.
As $\dot{\phi}$ and $\dot{A}_a^i$ are the only time derivatives in Eq.~\eqref{Lha}, they form canonical variables through a symplectic structure with $\pi_\phi$ and $E^{a i}/\beta$ as the respective canonical momenta.

Since our aim is to construct a conformally extended Hamiltonian formulation of gravitation with first class constraints as symmetry generators, any second class constraints should be eliminated using variational field equations not of the canonical type. From the canonical analysis of the standard Holst action \cite{Holst1996}, it is clear that such second class constraints will also arise from the torsion-free condition in our generalized formulation. Therefore, we eliminate such second class constraints from  $\GG_a^i$ and $\HH_{ab\,IJ}$ using Eq.~\eqref{A2Gam} already derived as part of the variational field equations. After working through the remaining rather involved algebra, we arrive at first class constraints of anticipated forms which are in full agreement with their independent derivations via a conformally extended Arnowitt-Deser-Misner (ADM) formulism. (Contained in Supplementary Material Sec.~II  \cite{SuppM})
The resulting constructions allow concrete connections with established loop quantization techniques with considerably new and interesting conformal and scaling features.

In applying Eq.~\eqref{A2Gam}, one notices that the LC connection $\Gamma_{IJ}$ encodes the spin connection 1-forms
$
\Gamma_a^i
=
-
\epsilon_{ijk}
\Gamma_{a jk}/2
$
and extrinsic curvature 1-forms
$
{K}_{a i}
=
\Gamma_{a i 0}
$.
As a result, $A_a^i$ introduced above becomes the {\it conformal spin connection variable}
\begin{eqnarray}
A_a^i
&=&
\Gamma_a^i + \beta C_a^i
\label{AAai}
\end{eqnarray}
which is the sought after {\it conformal Ashtekar-Barbero connection variable},
in terms of the {\it conformal extrinsic curvature} 1-forms
\begin{eqnarray}
C_a^i
&=&
\phi^2
(K_a^i + \phi_{a i 0})
\label{Kia}
\end{eqnarray}
with the associated curvature 2-forms
$
F_{ab}^i
=
2\, \partial_{[a} A_{b]}^i + \epsilon_{ijk} A_a^j A_b^k
$.
The scalar momentum then becomes
\begin{eqnarray}
\pi_\phi
&=&
-
\frac{6\,\sqrt{E}}{N}\,(\dot{\phi}-N^a\phi_{,a})
+
2\,\sqrt{E}\,\phi\,K
\label{piphi}
\end{eqnarray}
where $K=-K_a^i e^{a}{}_i$, is the trace of the extrinsic curvature tensor.

A conformal redundancy among $C_a^i$ and $\pi_\phi$ gives rise to a new {\it conformal constraint}
\begin{eqnarray}
\CC
&=&
\phi\,\pi_\phi
+
2\,C_a^i E^a_i
\label{confcons}
\end{eqnarray}
which is required to vanish weakly.
It is in fact the canonical generator of the conformal frame transformations
\begin{subequations}
\begin{eqnarray}
\hspace{-10pt}
\phi
&\to&
\Omega^{-1}\phi
,\;\;\;\;
\pi_\phi
\to
\Omega\,\pi_\phi
,\;\;\;\;
E^a_i
\to
\Omega^2 E^a_i
,
\label{cnftrans1}
\ppp
\hspace{-20pt}
A_a^i
&\to&
A_a^i
+
\beta
(\Omega^{-2}-1)
C_a^i
-
\frac{1}{4}\epsilon_{ijk}
E_a^j E^b_k\,
(\ln\Omega)_{,b}
\label{cnftrans2}
\end{eqnarray}
\label{cnftrans}
\bppp
\end{subequations}
for any positive {\it function} $\Omega(x)$.

The transformation relation for $A_a^i$ above originates from the conformal transformation of the LC spin connection
$\Gamma_a^i \to \Gamma_a^i + \text{[last term in Eq.~\eqref{cnftrans2}]}$
and the conformally {\it covariant} change of the conformal extrinsic curvature
$C_a^i \to \Omega^{-2} C_a^i$ under Eq.~\eqref{cnftrans1}, a crucial property ensuring the overall conformal covariance of our new formalism not shared by the standard extrinsic curvature $K_a^i$.

After some calculations, the expressions involving $\GG_a^i$, $\HH_{ab\,i0}$,  and $\HH_{ab\,ij}$ in Eq.~\eqref{Lha} along with the zero-torsion condition Eq.~\eqref{A2Gam} significantly simplify to further constraints as below.
First, the term $E^a_i \GG_a^i$,
up to a total divergence, becomes
$-M_i \GG_i$,
where
$M_i$
are certain functions of the nondynamical
$\Gamma_{t j k}$
and
$\Gamma_{t i 0}$,
hence acting as a Lagrangian multiplier. Here,
\begin{eqnarray}
\GG_i
&=&
D_a \, E^a_i/\beta
\label{gausscons}
\end{eqnarray}
takes the standard form of the {\it Gauss constraint}, as the canonical generator of the spin gauge transformations, now using the spin covariant derivative $D_a$ with respect to the conformal spin connection $A_a{}^i$ given by Eq.~\eqref{AAai}.

Up to adding a quantity proportional to the Gauss constraint $\GG_i$, the term
$E^b{}^i \HH_{a b\,i 0}$
reduces to
\begin{eqnarray}
\HH_a
&=&
F_{ab}^i\, E^b_i/\beta
+
\phi_{,a} \pi_\phi
\label{diffeomcons}
\end{eqnarray}
of the standard form of {\it diffeomorphism constraint} as the canonical generator of the spatial diffeomorphisms. Note also, $F_{ab}^i$ above uses the conformal spin connection \eqref{AAai} and $\phi$ is the conformal scalar field.


The remaining term
$
-1/(2\sqrt{E}\,)
E^a_i  E^b_j\, \HH_{a b\,i j}
$
in Eq.~\eqref{Lha} simplifies,
up to adding a quantity proportional to the covariant derivative of the Gauss constraint $\nabla_a\GG_i$,
to be
\begin{eqnarray}
\HH
&=&
\phi^2\HH_\E
-
\frac{1}{\sqrt{E}\,\phi^2}
\big(
1 + \beta^2\phi^4
\big)
C_{[a}^{i} C_{b]}^{j}
E^a_i E^b_j
\nppp
&&
+
\sqrt{E}\big(
2\phi\Delta\phi
-
\phi_{,a}\phi^{,a}
\big)
\label{hamcons}
\end{eqnarray}
where $\Delta$ is the Laplace-Beltrami operator of $h_{ab}$,
which is our new {\it Hamiltonian constraint}. Apart from its subexpression
\begin{eqnarray}
\HH_\E
&=&
\frac{1}{2\sqrt{E}}\,\epsilon_{i j k}
F_{ab}^k\,
E^a_i E^b_j
\label{HE}
\end{eqnarray}
of the standard form of the {\it Euclidean Hamiltonian constraint}, it is modified from the standard Hamiltonian constraint with additional presence of the conformal scalar field $\phi$ and the replacement of the extrinsic curvature by conformal extrinsic curvature $K_a^i \to C_a^i$, as well as $F_{ab}^i$ using the conformal spin connection \eqref{AAai}.

Using Eqs.~\eqref{gausscons}, \eqref{diffeomcons}, and \eqref{hamcons}, we obtain from Eq.~\eqref{Lha} the totally constrained Hamiltonian density
%
\begin{eqnarray}
\HHH
&=&
M_i \GG_i
+
N^a \HH_{a}
+
N \HH .
\label{Lhacons}
\end{eqnarray}
Given their structures, it is clear from previous related studies that constraints $\CC, \GG_i, \HH_a$ and $\HH$ are first class
\cite{Veraguth2017, Wang2005a, Wang2005b, Wang2006a}.

As inherited from the conformally extended Holst action \eqref{S}, physical observables are quantities invariant under,
in addition to rotations and diffeomorphisms,
conformal transformations Eqs.~\eqref{cnftrans}.
An important example is the conformally invariant triad
$\ovr{E}^a_i=\phi^2E^a_i$ which plays the role of the physical triad and gives rise to physically measurable geometric quantities. Specifically, the physical area operator on a surface in $\Sigma$ is given by \cite{Rovelli1995}
\newpage
\begin{eqnarray}
\nonumber\\[-25pt]
A_S
&=&
\int_S |\ovr{E}|
=
\int_S \phi^2 |E|
\label{AS}
\end{eqnarray}
where $|\ovr{E}|$ ($|E|$) is the norm of the 2-form as the pullback of $\ovr{E}^a_i$ (${E}^a_i$) to $\Sigma$.
Upon quantization, any discrete spectrum from $|E|$ can be blurred by the dense eigenvalues of $\phi$, resulting in the absence of definitive area gaps in either standard loop quantization \cite{Rovelli1995} or other schemes, e.g. the Fock representation of null surface area spectrum recently studied in Ref.~\cite{Wieland2017}.

This stark contrast also begs the question on the Immirzi parameter, which controls the size of e.g. the area gaps in standard LQG. What is the physical effect of our conformal Immirzi parameter $\beta$ if there are no definitive area gaps? It turns out that there is no preferred value of $\beta$ for the new theory in as much as no privileged conformal frames.

Indeed, the parameter $\beta$ that appears in all the constraints and multipliers contained in the Hamiltonian density \eqref{Lhacons} can be completely transformed away, leading simply to
\begin{eqnarray}
\HHH
\text{\raisebox{-1pt}{
$\stackrel{\Omega=\sqrt{\beta}}{-\!\!\!-\!\!\!-\!\!\!\longrightarrow}$ }}
\HHH|_{\beta=1}
\label{HHHtrans}
\end{eqnarray}
using the ``normalized'' conformal Ashtekar-Barbero variable
$A_a^i = \Gamma_a^i + C_a^i$.

\section{Conformal loop quantization}
\label{sec:d}

While our connection degrees of freedom are amenable to loop quantization,
since we aim to preserve the conformal invariance at he quantum level, it is natural to quantize $\phi$ in the field representation. Following Ref.~\cite{Veraguth2017}, we consider the kinematic Hilbert space
\begin{eqnarray}
\HH_\text{kin}=\HH_\text{SF}\otimes\HH_\text{SN}
\end{eqnarray}
as a product of that for scalar fields (SF) and spin networks (SN),
like LQG coupled to a scalar field in Ref.~\cite{Lewandowski2016}. The quantum states are therefore expressed as superpositions of
\begin{eqnarray}
\Psi[\phi,A]
=
\Psi[\phi]\otimes\psi(h_{e_1}[A],\ldots,h_{e_n}[A])
\end{eqnarray}
using the cylindrical functions of $A_a^i$,
with holonomies $h_{e_1}[A],\ldots,h_{e_n}[A]$ over edges
$e_1,\ldots,e_n$.
In particular, the implementation of Eqs.~\eqref{cnftrans} as quantum canonical transformation amounts to the loop quantization of the conformal constraint \eqref{confcons}.
After mapping $K_a^i \to C_a^i$ the definitions of our connection variables \eqref{AAai} and the Euclidean Hamiltonian constraint \eqref{HE} are identical to standard definitions, and so the nontrivial quantization of the mean extrinsic curvature term in Eq.~\eqref{confcons} follows from Ref.~\cite{Thiemann1996} as
\begin{eqnarray}
C_a^i E^a_i(x)
&=&
\frac{1}{i\hbar}\,
[
\HH_\E(x) , V_\Sigma
]
\end{eqnarray}
where $V_\Sigma$ is the volume operator on the spatial slice $\Sigma$ with respect to $E^a_i$.
Since the resulting loop quantization is independent of $\beta$, we can put $\beta=1$. Together with all quantized constraints, the physical state $\Psi$ is therefore described by the solutions of the quantum constraint equations
$(\HH,\,\HH_a,\GG_i)\, \Psi = 0$
analogous to standard LQG, augmented with the new quantum conformal constraint equation
\begin{eqnarray}
\CC \Psi = 0
\label{Cpsix}
\end{eqnarray}
which is to be implemented through e.g. conformal group averaging along with diffeomorphisms to form conformorphism invariant states \cite{Wang2005b}.

If the constraint \eqref{Cpsix} is to be preserved when matter coupling is also included, then matter must be conformally invariant too. One possibility is to consider conformally invariant scalar-tensor (ST) theory of gravity
\cite{Wagoner1970, Wang2013}
which is equivalent to GR at least classically.
From a minimal coupling point of view, if we assume that the metric coupling with matter is independent of the scalar field $\phi$, then conformal invariance can still hold for the general covariant SM with a conformally coupled Higgs field, following the {\it principle of conformal invariance} \cite{Bekenstein1980}.

\section{Scale-invariant gravity-scalar-matter system}
\label{sec:e}

If one does not insist on the full conformal invariance, then
Eq.~\eqref{HHHtrans}
can still be achieved as a {\it scale-invariant} system
with physical states $\Psi$ satisfying the global conformal constraint, i.e. scaling constraint
\begin{eqnarray}
\int\CC\,\dd^3x\; \Psi
&=&
0
\label{Cpsi}
\end{eqnarray}
instead of the local constraint \eqref{Cpsix}.


Indeed, motivated by the fact that only scale invariance, not the full conformal invariance, is required to achieve an Immirzi parameter-free description, let us now consider the following scale-invariant Lagrangian density
\begin{eqnarray}
\LLL
&=&
\LLL_\G[g_{\mu\nu},\phi]
+
\LLL_\S[g_{\mu\nu},\phi,\A_\mu]
\nppp&&
+
\LLL_\M[g_{\mu\nu},\phi,\A_\mu,\psi]
\label{LLLtot}
\end{eqnarray}
for a coupled system with gravity (G), scalar (S), and matter (M), based on a naturally generalized SM  approach with scale invariance.

Given the equivalence between the first order Holst formalism and second order ADM-type formulism demonstrated above, let us proceed with the ADM-type canonical analysis below. (Technical details are available in Supplementary Material Sec.~III \cite{SuppM}).
In the Lagrangian density~\eqref{LLLtot} above, we consider
\begin{eqnarray}
\LLL_\G
&=&
\frac12\,\Phi^2
\sqrt{-g}\,R
\label{LLLG}
\end{eqnarray}
where $R$ is the scalar curvature of $g_{\mu\nu}$ and
$
\Phi^2
=
\sum_{A}{\phi_A\phi^A}/{6k_{A}}
$.
Here $k_{A}$ are scalar-gravity coupling constants, with $k_A=1$ for any conformally coupled $\phi^A$.
Part of the index range for $A$ may be the Haar measure indices associated with a gauge group $G$.
%
%
In Eq.~\eqref{LLLtot}, the term
\begin{eqnarray}
\LLL_\S
&=&
-
\sqrt{-g}\,
\Big[
\frac{1}{2}\,
\DD_\mu\phi_A \DD^\mu\phi^A
+
V(\phi)
\Big]
\label{LLLS}
\end{eqnarray}
is the scalar Lagrangian density, where $V(\phi)$ is a homogeneously fourth order potential in $\phi^A$ \cite{Bekenstein1980} that may account for both mass and potential of the Higgs scalar, and $\DD_\mu$ is the covariant derivative using the sum of the LC connection and gauge connection $\A_\mu$.



The matter Lagrangian density in Eq.~\eqref{LLLtot} is
\begin{eqnarray}
\LLL_\M
&=&
\Re\,\sqrt{-g}\,
\psi^\dagger i\gamma^0
\big[
\gamma^I e^\mu{}_I
\DD_\mu
+
\mu(\phi)
\big]
\psi
\nonumber\\[5pt]
&&
-
\frac14\,\sqrt{-g}\,
\FF_{\mu\nu}^A\FF^{\mu\nu}_A
\label{LLLM}
\end{eqnarray}
which uses the curvature $\FF_{\mu\nu}$ of $\A_\mu$,
the constant Dirac matrices
$\gamma^I$
satisfying the Clifford algebra
$\gamma^{(I}\gamma^{J)} = \eta^{IJ}$
with the Hermiticity condition
$\gamma^I{}^\dag = \gamma^0\gamma^I\gamma^0$,
and a Yukawa coupling matrix
$\mu(\phi)$
homogeneous linearly in $\phi^A$ \cite{Bekenstein1980}.

Lagrangian density \eqref{LLLtot} has the property of being invariant under the following scale transformations:
\begin{eqnarray}
\begin{array}{c}
g_{\mu\nu}
\to
\Omega^2 g_{\mu\nu}
,\quad
\phi
\to
\Omega^{-1}\phi
,
\\[10pt]
\A_\mu
\to
\A_\mu
,\quad
\psi
\to
\Omega^{-3/2}\psi,
\end{array}
\label{scltrans}
\end{eqnarray}
for any positive {\it constant} $\Omega$, as a result of
$\LLL_\G$, $\LLL_\S$, and $\LLL_\M$ being similarly invariant.

The scaling symmetry implies the existence of a conserved Noether current~\cite{Banados2016},
which can be identified from the boundary terms of the {\it on-shell} variations of the Lagrangian density \eqref{LLLtot} under a scaling transformation with $\Omega=1+\varepsilon$ in Eq.~\eqref{scltrans} for an infinitesimal $\varepsilon$, denoted by
$\delta_\varepsilon \sqcup$.
To this end, first it can be obtained from Eq.~\eqref{LLLG} that
\begin{eqnarray*}
\delta_\varepsilon
\LLL_\G
&=&
\frac12\,
\sqrt{-g}\,
\nabla_\mu
\big(
\Phi^2
g^{\alpha\beta}\delta_\varepsilon\Gamma^\mu_{\alpha\beta}
-
\Phi^2
g^{\mu\beta}\delta_\varepsilon\Gamma^\alpha_{\alpha\beta}
\big)
=
0
\end{eqnarray*}
where $\nabla_\mu$ is the LC covariant derivative using $g_{\mu\nu}$,
on account of the invariance of the LC connection under scaling transformations, and hence $\delta_\varepsilon\Gamma^\mu_{\alpha\beta}=0$.
A nontrivial contribution is derived from Eq.~\eqref{LLLS} as follows
\begin{eqnarray*}
\nonumber\\[-3pt]
\delta_\varepsilon
\LLL_\S
&=&
\frac{\epsilon}{2}\,
\sqrt{-g}\, \nabla_\mu
\partial^\mu\phi^2
\end{eqnarray*}
where
$\phi^2 = \phi_A\phi^A$,
since $\delta_\varepsilon \phi=-\varepsilon \phi$.
There are no further relevant boundary terms resulting from varying gauge connections since they are invariant under the scale transformations, i.e. $\delta_\varepsilon \A_\mu=0$.
By varying the Lagrangian density $\eqref{LLLM}$, one finds that the scale transformations of spinors under Eq.~\eqref{scltrans} do not contribute to any boundary term either, as
$\delta_\varepsilon\LLL_\M=0$.
It therefore follows from the above discussions, that the on-shell variational relation $\delta_\epsilon\LLL=0$ gives rise to the conserved {\it Weyl current}
$\partial_\mu\phi^2$. See also related discussion in Ref.~\cite{Ferreira2017}.

To focus on the gravitational structure with multiple scalar fields, in the following we suppress the gauge field $\A_\mu$, spinor $\psi$, and the scalar potential $V(\phi)$, so that
$\DD_\mu\phi^A
=
\partial_\mu\phi^A$
and
$\LLL
=
\LLL_\G+\LLL_\S$.
The off-shell variations of $\LLL$ yields the canonical momentum for the metric
\begin{eqnarray}
p^{ab}
&=&
-
\frac{\sqrt{h}}{N}\,h^{ab}\Phi(\dot{\Phi}-N^c\Phi_{,c})
\nonumber\\[3pt]
&&
-
\frac{\Phi^2\sqrt{h}}{2}\,(K^{ab}-h^{ab}K)
\label{ppab}
\end{eqnarray}
and the canonical momentum for the scalars
\begin{eqnarray}
\pi_A
&=&
-
\frac{\sqrt{h}}{N}
\big[
\dot{\phi}_A
-
N^a \phi_{A,a}
\big]
+
\frac{\sqrt{h}}{3\kappa_A}\, K \phi_A .
\label{piphi1}
\end{eqnarray}

The conformal extrinsic curvature \eqref{Kia} and the conformal constraint \eqref{confcons} are generalized to be
$
C_a^i
=
\phi^2
(K_a^i + \Phi_{a i 0})
$
and
$
\CC
=
\phi^A \pi_{A}
+
2\,C_a^i E^a_i
$
respectively, where $\Phi_{a i 0} = \phi_{a i 0}|_{\phi\to\Phi}$ with the conformal Ashtekar-Barbero variable $A_a^i$ taking the same form as Eq.~\eqref{AAai}.
Other constraints also follow analogously.
For example, the Hamiltonian constraint for a set of nonconformal scalar fields $\phi_A$, with a common $k_A=k\neq1$, for all $A$ is found to be
\begin{eqnarray}
\HH
&=&
\Phi^2\HH_\E
-
\frac{1}{\sqrt{E}\,\Phi^2}
\big(
1 + \beta^2\Phi^4
\big)
C_{[a}^{i} C_{b]}^{j}
E^a_i E^b_j
\nonumber\\[3pt]
&& 
+
\sqrt{E}\,\Delta\Phi^2
-
\frac{\sqrt{E}}{2}
\phi_{A,a}\phi^{A,a}
-
\frac{1}{12k(k-1)\Phi^2\sqrt{E}}
\nonumber\\[3pt]
&&
\times\,
\Big[
k\,\CC^2
-
\phi^A \pi^B \pi_A \phi_B
+
\phi^A \phi^B \pi_A \pi_B
\Big]
\label{hhhc}
\end{eqnarray}
which has a generalized but similar structure as the Hamiltonian constraint \eqref{hamcons} and therefore is also amenable to the scalar-loop quantization described above.

The corresponding scale-invariant physical triad is given by
$\ovr{E}^a_i=\Phi^2E^a_i$ giving rise to physically measurable geometric quantities such as the area operator
\begin{eqnarray}
A_S
&=&
\int_S |\ovr{E}|
=
\int_S \Phi^2 |E|
\label{AS2}
\end{eqnarray}
analogous to Eq.~\eqref{AS}.
For a single scalar field, Hamiltonian constraint \eqref{hhhc} recovers the Hamiltonian constraint in Ref.~\cite{Veraguth2017}. The corresponding Hamiltonian density also takes the form of Eq.~\eqref{Lhacons}, which under a scaling transformation of the form of Eq.~\eqref{cnftrans} with $\Omega=\sqrt{\beta}$ also sets $\beta=1$ as in Eq.~\eqref{HHHtrans} as alluded to above.

\vspace{-10pt}
\section{Conclusion and discussion}

\vspace{-10pt}
In this paper, we have addressed a number of longstanding issues of significance in LQG related to the uniqueness of loop variables, discreteness of quantum geometry, and to a smaller extent, possible emergence of time, based on an important principle---{\it conformal invariance}.

In Sec.~\ref{sec:b}, we have derived conformally extended Ashtekar-Barbero variables leading to the new loop variables of gravity that are free from the Immirzi ambiguity by effectively absorbing an otherwise one-parameter family of variables into a conformal class. Although variables of the same form have previously appeared in Ref.~\cite{Veraguth2017}, where they were derived in a specific ST theory excluding GR and matter coupling, here we have demonstrated the conformal Ashtekar-Barbero variables further arise from a large general class of gravity theories covering GR, ST, dilatonic gravity. In addition, we show that our formalism is preserved for matter coupling with SM-type theories. This is due to their crucial conformal or scale invariance. In so doing, we have addressed an often raised question on whether LQG restricts matter coupling and have suggested the naturalness, if not uniqueness, of the SM-type matter that can be coupled to quantum gravity. Furthermore, instead of the more limited canonical formalism adopted in Ref.~\cite{Veraguth2017}, the new conformal Holst principle in this work paves the way for a more covariant spinform models of the new conformal LQG to be developed.

Since our conformal Immirzi parameter $\beta$ can be arbitrarily set to unity, from Eq.~\eqref{hamcons} and more generally Eq.~\eqref{hhhc}, it is readily seen that in the case of constant scalar fields, $\Phi^2$ acts like the conventional Immirzi parameter $\gamma$. Indeed, the area operator \eqref{AS2} assumes the standard LQG form with the linear dependence on the Immirzi parameter for $\Phi^2\to\gamma$.
How can we, then, reconcile with the apparent value $\gamma=\gamma_0\simeq0.274$ fixed from black hole entropies \cite{Perez2017}? One possibility to be investigated would be that, in a suitable conformal frame (at the kinematic level before conformorphism group averaging), $\Phi^2=\gamma_0$ might turn out to be thermodynamically favorable on the black hole horizon. However, such an effective Immirzi parameter $\Phi^2$ may dynamically vary e.g. in the early Universe \cite{Navascues2018}.

Our parameterless formalism can also be understood, in the terminology of BF gravity \cite{Celada2016}, as the equivalence between the starting gravitational action \eqref{S} and the following deformed BF action
\begin{eqnarray*}
S
&=&
-\frac12\,\int
\Big[
B^{IJ}\wedge F_{IJ}
-
\ovr{\star\, B^{IJ} \wedge  F_{IJ}}\,
\Big]
\end{eqnarray*}
with 2-forms $B^{IJ} = e^I \wedge e^J$ and without coupling constants. This is achieved by, without loss of generality, setting the secondary conformal factor $\theta=1/\sqrt{\beta}=1$ as explained in Sec.~\ref{sec:c}. The overline in the above action denotes a conformal transform with respect to $\Phi$ in general as described in Sec.~\ref{sec:e}.

In the case of scale-invariant dilaton systems in Sec.~\ref{sec:e}, our new framework is relevant for the quantum gravitational temporal evolution of the early Universe~\cite{Brizuela2016}, particularly in relation to the conserved Weyl current giving rise to a natural cosmological time \cite{Ferreira2017, Ferreira2018}. In this respect, a lack of definitive area gaps might lead to a radically different conclusion to the (non)existence of the big bang or big bounce from standard loop quantum cosmology (LQC).
The Weyl current $\partial_\mu\Phi^2$ suggests $\Phi^2$  may be a ``fuzzy''  harmonic time for the ultimate physical wave functional
$
\Psi[\psi, \A_a, A_{a}{}^{i}, \hat{\phi}^A\, ; \Phi^2]
$
in terms of the normalized scalar fields
$
\hat{\phi}^A
=
\phi^A/{\sqrt{\Phi^2}}\,
$.
Although here we have separated out $\Phi^2$, it need not be a deparametrized functional time~\cite{York1971, Isham1992, Rovelli1994, Husain2012}. Instead, we could start with a timeless Wheeler-DeWitt type equation and only recover $\Phi^2$ as an emergent time in the quantum regime as in LQC models \cite{Agullo2016} or
in the (semi)classical limit as in Refs.~\cite{Ferreira2017, Ferreira2018}. Further research is required to address the above open questions and we hope to report on related progress in future work.

\begin{acknowledgments}

C.~W. wishes to thank C. J. Isham and J. W. York for encouragement in pursuing the conformal properties of (loop) quantum gravity. The referee's constructive criticism has been very helpful in clarifying the presentation of results and their context. For financial support, C.~W. and D.~R. are grateful to EPSRC, STFC, and Cruickshank Trust (UK) and CNPq (Brazil).

\end{acknowledgments}

\end{document}


\title{\Large\bf Supplementary Material to ``Closing the gaps in quantum space and time: Conformally augmented gauge structure of gravitation''}




\author{Charles H.-T. Wang\footnote{c.wang@abdn.ac.uk}\\[5pt]
{\normalsize Department of Physics, University of Aberdeen, King's College, Aberdeen AB24 3UE, United Kingdom}}



\date{}

\maketitle

\begin{quote}
This Supplementary Material contains some detailed mathematical constructions and derivations of the theoretical results presented in the Main Text. It is organized into the following three sections:
\\

\S\ref{sec:1} Conformally extended Holst action principle for general relativity.
\\

\S\ref{sec:2} Generalized ADM formalism with conformal Ashtekar-Barbero variables.
\\

\S\ref{sec:3} Conformally augmented gauge theory of scale-invariant dilaton gravitation.

\end{quote}

\renewcommand{\S}{\text{S}}

\section{Conformally extended Holst action principle for general relativity}
\label{sec:1}

\subsection{Conformal changes of the metric and connection}

Consider a 4-dimensional Lorentzian manifold with a (covariant) metric tensor
\begin{eqnarray}
\g =  \eta_{IJ} e^I \otimes e^J
\label{gab}
\end{eqnarray}
using the orthonormal co-frame
$\{ e^I \}, I=0,1,2,3$,
with the Minkowski components
\begin{eqnarray}
\eta_{IJ} = \diag(-1, 1, 1, 1).
\label{mink}
\end{eqnarray}

The corresponding contravariant metric is given by
\begin{eqnarray}
\tilde{\g} =  \eta^{IJ} \tilde{e}_I \otimes \tilde{e}_J
\label{ginvab}
\end{eqnarray}
using the orthonormal frame $\{\tilde{e}_I \}$ so that they contract to the Kronecker delta
\begin{eqnarray}
e^I(\tilde{e}_J) = \delta^I_J.
\label{ee}
\end{eqnarray}

Consider the general antisymmetric connection 1-forms $A_{IJ}$.
Its curvature 2-forms $F_{IJ}$ are given by
\begin{eqnarray}
F_{IJ}
&=&
\dd A_{IJ} + A_{IK} \wedge A^K{}_J.
\label{Rab}
\end{eqnarray}

Using the covariant exterior derivative $\DDD$ associated with $A_{IJ}$, the related torsion 2-forms are given by
\begin{eqnarray}
T^{I}
&=&
\DDD e^I
=
\dd e^I + A^I{}_J \wedge e^J
\label{tor}
\end{eqnarray}
satisfying the first Bianchi identity
\begin{eqnarray}
\DDD T^{I}
&=&
\dd T^I + A^I{}_J \wedge T^J
=
F^I{}_J \wedge e^J.
\label{DTI}
\end{eqnarray}

Substituting \eqref{Rab} into \eqref{DTI}, we obtain the relation
\begin{eqnarray}
\DDD T_{I}
&=&
F_I{}_J \wedge e^J
=
\dd A_{IJ}  \wedge e^J + A_{IK} \wedge A^K{}_J \wedge e^J.
\label{DTI2}
\end{eqnarray}
It follows that $\DDD\, \eta_{IJ}=0$ and so the antisymmetric $A_{IJ}$ is metric compatible.

The Levi-Civita (LC) connection 1-forms $\Gamma_{IJ}$ are given in terms of the tetrad by
\begin{eqnarray}
\Gamma_{IJ}
&=&
\frac12 ( e^K \ii_{I}\ii_{J} \dd e_K + \ii_{J} \dd e_I - \ii_{I} \dd e_J )
\label{wab}
\end{eqnarray}
(see e.g. \cite{Tucker1987}) where
$\ii_{I} := \ii_{\tilde{e}_I}$
and tetrad indices $I,J,\cdots$ are raised or lowered using $ \eta^{IJ}$ and $ \eta_{IJ}$.
We will also use the shorthand
$f_{,I}=\ii_I \dd f$
with
$f^{,I}=\ii^I \dd f=\eta^{IJ}\ii_J \dd f$
for any scalar function $f$.

It is well-known that the LC connection is the unique metric compatible and torsion free connection, satisfying in particular
\begin{eqnarray}
\DDD\, e^I
=
0
\label{torfr}
\end{eqnarray}
using the covariant exterior derivative associated with $A_{IJ}=\Gamma_{IJ}$.
Consequently, Eqs. \eqref{Rab} and \eqref{DTI2} yield the following two relations
\begin{gather}
\dd e^I + \Gamma^I{}_J \wedge e^J
=
0
\label{Dea}
\ppp
\dd \Gamma_{IJ}  \wedge e^J + \Gamma_{IK} \wedge \Gamma^K{}_J \wedge e^J
=
0.
\label{DGa}
\end{gather}

Under a conformal transformation using a positive scalar function $\phi$, denoted with an over-bar,
\begin{subequations}
\begin{eqnarray}
\ovr{\g} &=& \phi^2 {\g}
\label{gbar1}
\ppp
\label{gbar2}
\ovr{\tilde{\g}} &=& \phi^{-2} \tilde{\g}
\end{eqnarray}
\bppp
\end{subequations}
the covariant \eqref{gab} and contravariant \eqref{ginvab} metric  become
\begin{subequations}
\begin{eqnarray}
\ovr{\g}
&=&
\eta_{IJ}\, \ovr{e}^I \otimes \ovr{e}^J
\ppp
\ovr{\tilde{\g}}
&=&
\eta^{IJ}\, \ovr{\tilde{e}}_I \otimes \ovr{\tilde{e}}_J
\end{eqnarray}
\bppp
\end{subequations}
in terms of the corresponding conformally transformed tetrad as follows:
\begin{subequations}
\begin{eqnarray}
\ovr{e}^I
&=&
\phi\, {e}^I
\label{be}
\ppp
\ovr{\tilde{e}}_I
&=&
\phi^{-1}\tilde{e}_I.
\end{eqnarray}
\bppp
\end{subequations}

The induced conformal transformation of the LC connection $\Gamma_{IJ}$ follows from \eqref{wab} with $e^I\to\ovr{e}^I$ using \eqref{be} to be
\begin{eqnarray}
\ovr{\Gamma}_{IJ}
&=&
\Gamma_{IJ} + \phi_{IJ}
\label{bwablc}
\end{eqnarray}
where we have introduced
\begin{eqnarray}
\phi_{IJ}
&=&
\phi^{-1}( e_I \ii_J \dd\phi - e_J \ii_I \dd\phi )
\nppp
&=&
-2\phi^{-1}\phi_{,[I}  e_{J]}.
\label{psiab}
\end{eqnarray}

Applying the conformal transformation to \eqref{torfr}, we see that $\ovr{\Gamma}_{IJ}$ is also torsion free and hence
\begin{eqnarray}
0
&=&
\DDD\, \ovr{e}^I
\nppp
&=:&
\dd \ovr{e}^I + \ovr{\Gamma}^I{}_J \wedge \ovr{e}^J
\nppp
&=&
\phi\,\DDD\,{e}^I
+
\dd\phi\wedge \dd e^I + \phi\,\phi^I{}_J \wedge e^J.
\label{Dea2}
\end{eqnarray}
The first term above vanishes using \eqref{Dea}.
Because $\phi>0$, the 2nd term above then yields the identity
\begin{eqnarray}
\dd\phi\wedge \dd e^I + \phi\,\phi^I{}_J \wedge e^J
= 0.
\label{Deaid}
\end{eqnarray}

Using \eqref{Deaid} and denoting by $\ovr{A}_{IJ}$ the conformally transformed connection 1-forms $A_{IJ}$ with the associated covariant exterior derivative $\ovr\DDD$, we see that the following conformally covariant relation
\begin{eqnarray}
\ovr{\DDD}\, \ovr{e}^I
&=&
\phi\,\DDD {e}^I
\label{Deacov}
\end{eqnarray}
is satisfied if $\ovr{A}_{IJ}$ is given by
\begin{eqnarray}
\ovr{A}_{IJ}
&=&
A_{IJ} + \phi_{IJ}.
\label{bwab}
\end{eqnarray}

This relation will therefore be taken as the conformal transformation of $A_{IJ}$. In particular, if the torsion of $A_{IJ}$ vanishes, the above reduces to \eqref{psiab}.

Instead of $\phi$, we can use a different function $\phi\to\theta$ with $\ovr{(\cdots)}\to\und{(\cdots)}$ and \eqref{psiab} becoming
\begin{eqnarray}
\theta_{IJ}
&=&
\theta^{-1}( e_I \ii_J \dd\theta - e_J \ii_I \dd\theta )
\nppp
&=&
\theta^{-1}(\theta_{,J}  e_I - \theta_{,I}  e_J)
\nppp
&=&
\theta^{-1}(\dd\theta(\tilde{e}_J)  e_I - \dd\theta(\tilde{e}_I)  e_J ).
\label{ThetaIJ}
\end{eqnarray}

\subsection{Conformal changes of the curvature forms}

The curvature 2-forms \eqref{Rab} then transform as follows
\begin{eqnarray}
\ovr{F}_{IJ}
&=&
\dd \ovr{A}_{IJ} + \ovr{A}_{IK} \wedge \ovr{A}^K{}_J
\nppp
&=&
F_{IJ} + \DDD\phi_{IJ} + \phi_{IK} \wedge \phi^K{}_J.
\label{FP}
\end{eqnarray}

Likewise, we have
\begin{eqnarray}
\und{F}_{IJ}
&=&
\dd \und{A}_{IJ} + \und{A}_{IK} \wedge \und{A}^K{}_J
\nppp
&=&
F_{IJ} + \DDD\theta_{IJ} + \theta_{IK} \wedge \theta^K{}_J.
\label{FQ}
\end{eqnarray}

The second and last terms of \eqref{FP} can be evaluated to yield
\begin{eqnarray*}
\DDD\phi_{IJ}
&=&
\big[
\dd(\phi^{-1})\phi_{,J}
+
\phi^{-1} \DDD \phi_{,J}
\big] \wedge e_I
-
(I \leftrightarrow J)
\nppp
&=&
\frac12\phi^{-2}\big[
\phi_{,J}\phi_{,L} \eta_{IK}
-
\phi_{,J}\phi_{,K} \eta_{IL}
+
\phi_{,I}\phi_{,K} \eta_{JL}
-
\phi_{,I}\phi_{,L} \eta_{JK}
\big]\, e^K\wedge e^L
\ppp
&&
+\frac12\phi^{-1} \big[
\phi_{;JK} \eta_{IL}
-
\phi_{;JL} \eta_{IK}
+
\phi_{;IL} \eta_{JK}
-
\phi_{;IK} \eta_{JL}
\big]\, e^K\wedge e^L
\end{eqnarray*}
and
\begin{eqnarray*}
\phi_{IK} \wedge \phi^K{}_J
&=&
\phi^{-2}
\big[
(\ii_J \dd\phi)(\ii^M \dd\phi) \eta_{IK}\eta_{de}
+ (\ii_I \dd\phi)(\ii^M \dd\phi) \eta_{ec}\eta_{JL}
- (\ii_M \dd\phi)(\ii^M \dd\phi) \eta_{IK}\eta_{JL}
\big]\,e^K \wedge e^L.
\end{eqnarray*}

Therefore the conformal transformation of the curvature 2-forms \eqref{FP} becomes
\begin{eqnarray}
\ovr{F}_{IJ}
&=&
F_{IJ} +
\frac12\phi^{-1} [
\phi_{;JK} \eta_{IL}
-
\phi_{;JL} \eta_{IK}
+
\phi_{;IL} \eta_{JK}
-
\phi_{;IK} \eta_{JL}
]\, e^K\wedge e^L
\nppp
&&
+
\frac12\phi^{-2}
[
\phi_{,M}\phi^{,M} \eta_{IL}\eta_{JK}
- \phi_{,M}\phi^{,M} \eta_{IK}\eta_{JL}
]\,e^K \wedge e^L
\nppp
&&
+\phi^{-2}[
\phi_{,J}\phi_{,L} \eta_{IK}
- \phi_{,J}\phi_{,K} \eta_{IL}
+ \phi_{,I}\phi_{,K} \eta_{JL}
- \phi_{,I}\phi_{,L} \eta_{JK}
]\,e^K \wedge e^L.
\label{BRab1}
\end{eqnarray}

Using \eqref{FP} we have
\begin{eqnarray}
\ovr{F_{IJ} \wedge \star (e^I\wedge e^J)}
&=&
\phi^{2}{F}_{IJ} \wedge \star ({e}^I \wedge {e}^J)
+
\phi^{2} \DDD\phi_{IJ} \wedge \star ({e}^I \wedge {e}^J)
+
\phi^{2} \phi_{IK} \wedge \phi^K{}_J \wedge \star ({e}^I \wedge {e}^J).
\label{Ree}
\end{eqnarray}

Using \eqref{be}, \eqref{psiab}, and \eqref{bwab}, we can calculate that
\begin{subequations}
\begin{eqnarray}
\ovr{\DDD\star (e^I\wedge e^J)}
&=&
\phi^{2} \DDD\star (e^I\wedge e^J)
+
2\phi \phi_{,K}e^K\wedge \star (e^I\wedge e^J)
\nppp&&
-
\phi\,\phi_{,K}  e^I \wedge \star (e^J\wedge e^K)
+3 \phi\,\phi^{,I} \star e^J
- \phi\,\phi_{,K} e^J \wedge \star (e^K\wedge e^I)
-3 \phi\,\phi^{,J} \star e^I
\nppp
&=&
\phi^{2} \DDD\star (e^I\wedge e^J)
\label{Dsee}
\ppp
\ovr{\DDD (e^I\wedge e^J)}
&=&
\phi^{2} \DDD(e^I\wedge e^J)
+
2\phi \phi_{,K}e^K\wedge e^I\wedge e^J
\nppp&&
-
\phi\,\phi_{,K}  e^I \wedge e^J\wedge e^K
- \phi\,\phi_{,K} e^J \wedge e^K\wedge e^I
\nppp
&=&
\phi^{2} \DDD(e^I\wedge e^J).
\label{Dseee}
\end{eqnarray}
\label{Dseeee}
\bppp
\end{subequations}

Relations \eqref{Dseeee} also follow more directly from \eqref{Deacov}.

\subsection{Variational relations}

In what follows, we denote the variation with respect to the scalar functions
$\phi$ and $\theta$, tetrad $e^I$, and connection $A_{IJ}$
by
$\delta_\phi$, $\delta_\theta$, $\delta_e$, and $\delta_A$
respectively.

Varying  \eqref{ee} yields
\begin{eqnarray}
\omega(\delta\tilde{e}_J)
=
-\delta{e}^I(\tilde{e}_J)\,\omega_I
\label{wee}
\end{eqnarray}
for any 1-form $\omega$.

Using \eqref{psiab} and \eqref{wee}, we obtain the following variation with respect to $e$
\begin{eqnarray}
\delta_e\phi_{IJ}
&=&
2\phi^{-1}(\dd\phi)_{[J} \delta e_{I]}
-
2\phi^{-1}(\dd\phi)_K (\delta{e}^K)_{[J} e_{I]}.
\label{psiabde}
\end{eqnarray}

Eq. \eqref{Rab}  gives the variation
\begin{eqnarray}
\delta F_{IJ}
&=&
\dd \delta A_{IJ}
-
A^K{}_{I} \wedge \delta A_{KJ}
-
A^K{}_J \wedge \delta A_{IK}
\nppp
&=&
\DDD \delta A_{IJ}
\label{dRab}
\end{eqnarray}

From \eqref{bwab} we see that
\begin{eqnarray}
\delta\ovr{A}_{IJ}
&=&
\delta A_{IJ} + \delta\phi_{IJ}.
\label{dbwab}
\end{eqnarray}

Similar to \eqref{dRab}, from  \eqref{dbwab} and \eqref{bwab} we obtain the variation
\begin{eqnarray}
\delta \ovr{F}_{IJ}
&=&
\dd \delta {A}_{IJ}
+ \delta {A}_{IK} \wedge \ovr{A}^K{}_J
- \delta {A}_{JK} \wedge  \ovr{A}^K{}_{I}
\nppp&&
+ \dd \delta \phi_{IJ}
+ \delta \phi_{IK} \wedge \ovr{A}^K{}_J
- \delta \phi_{JK} \wedge  \ovr{A}^K{}_{I}
\label{dARab}
\end{eqnarray}

Eq. \eqref{psiab} yields the following variation with respect to $\phi$
\begin{eqnarray}
\delta_\phi\phi_{IJ}
&=&
-\delta\phi\,\phi^{-1}\phi_{IJ}
+
\phi^{-1}( e_I \ii_J \dd\delta\phi - e_J \ii_I \dd\delta\phi ).
\label{dpsiab}
\end{eqnarray}

%
%

We have the relation
\begin{eqnarray}
\delta\,\ovr{\star (e^I\wedge e^J)}
&=&
2\delta\phi\,\phi \star (e^I\wedge e^J)
+
\phi^{2} \delta\star (e^I\wedge e^J)
\label{delsee}
\end{eqnarray}
where
\begin{eqnarray}
\delta\star (e^I\wedge e^J)
&=&
\delta e_K \wedge\star (e^I\wedge e^J\wedge e^K).
\label{delsee2}
\end{eqnarray}

Using  \eqref{be} and \eqref{FP} with $\phi\to\theta$ we have
\begin{eqnarray}
\und{F_{IJ} \wedge (e^I\wedge e^J)}
&=&
\theta^{2}{F}_{IJ} \wedge ({e}^I \wedge {e}^J)
+
\theta^{2} \DDD\theta_{IJ} \wedge ({e}^I \wedge {e}^J)
+
\theta^{2} \theta_{IK} \wedge \theta^K{}_J \wedge ({e}^I \wedge {e}^J).
\label{Reeh}
\end{eqnarray}

%
%

\subsection{Conformally extended Holst action}

Let us consider the action
\begin{eqnarray}
S
&=&
\frac1{2}\int
\Big[
\ovr{F_{IJ} \wedge \star (e^I\wedge e^J)}
-
\und{F_{IJ} \wedge e^I \wedge e^J}
\Big].
\label{S}
\end{eqnarray}

The first contribution in \eqref{S} is a conformally transformed Hilbert-Palatini (HP) term with $\g\to\ovr{\g}=\phi^2 \g$ whereas the second contribution in \eqref{S} is a conformally transformed Holst term  with $\g\to\und{\g}=\theta^2 \g$. Note that an Immirzi-like constant coefficient of the Holst term here can be absorbed into the function $\theta$.

\subsection{Variations with respect to the connection $A_{IJ}$}

To proceed, we do the $A$-variations first, which will result in the torsion free condition as follows.

From \eqref{dARab} we have the variation with respect to $A_{IJ}$ as follows
\begin{eqnarray}
\delta_A[\ovr{F_{IJ} \wedge \star (e^I\wedge e^J)}]
&=&
(\dd \delta {A}_{IJ}
+ \delta {A}_{IK} \wedge \ovr{A}^K{}_J
- \delta {A}_{JK} \wedge  \ovr{A}^K{}_{I})
\wedge \ovr{\star (e^I\wedge e^J)}
\nppp
&\deq&
\delta {A}_{IJ}
\wedge
\ovr{\DDD\star (e^I\wedge e^J)}
\label{dReeA}
\end{eqnarray}
where ``$\,\deq\,$'' signifies an equality modulo a total divergence.

Substituting \eqref{Dsee} into \eqref{dReeA} we see that
\begin{eqnarray}
\delta_A[\ovr{F_{IJ} \wedge \star (e^I\wedge e^J)}]
&\deq&
\delta {A}_{IJ}
\wedge
\phi^{2} \DDD\star (e^I\wedge e^J).
\label{dReeA2}
\end{eqnarray}

Consider the following variation with respect to $A_{IJ}$.
\begin{eqnarray}
\delta_A[\und{F_{IJ} \wedge (e^I\wedge e^J)}]
&=&
(\dd \delta {A}_{IJ}
+ \delta {A}_{IK} \wedge \und{A}^K{}_J
- \delta {A}_{JK} \wedge  \und{A}^K{}_{I})
\wedge \und{(e^I\wedge e^J)}
\nppp
&\deq&
\delta {A}_{IJ}
\wedge
\und{\DDD(e^I\wedge e^J)}.
\label{dReeeA}
\end{eqnarray}

Substituting \eqref{Dseee} into \eqref{dReeeA} we see that
\begin{eqnarray}
\delta_A[\und{F_{IJ} \wedge (e^I\wedge e^J)}]
&\deq&
\delta {A}_{IJ}
\wedge
\theta^{2} \DDD(e^I\wedge e^J).
\label{dReeeA3}
\end{eqnarray}

From \eqref{dReeA2} and \eqref{dReeeA3} we see that the $A$-variational equation of the action \eqref{S} i.e.
\begin{subequations}
\begin{eqnarray}
\delta_A[\ovr{F_{IJ} \wedge \star (e^I\wedge e^J)}
+
\und{F_{IJ} \wedge (e^I\wedge e^J)}]
=0
\label{dRecnd}
\end{eqnarray}
for any functions $\phi$ and $\theta$ yields
\begin{eqnarray}
\DDD \star ({e}^I \wedge {e}^J)
&=& 0
\label{dss1}
\ppp
\DDD  ({e}^I \wedge {e}^J)
&=& 0
\label{dss2}
\end{eqnarray}
\label{dss12}
\bppp
\end{subequations}
which are equivalent to the torsion-free condition \eqref{torfr}, requiring the connection $A_{IJ}$ to be LC.

\subsection{Variations with respect to the scalars $\phi$ and $\theta$}

From \eqref{dARab} and \eqref{dpsiab} we have the variation with respect to $\phi$ as follows
\begin{eqnarray}
\delta_\phi[\ovr{F_{IJ} \wedge \star (e^I\wedge e^J)}]
&=&
(\dd \delta_\phi {\phi}_{IJ}
+ \delta_\phi {\phi}_{IK} \wedge \ovr{A}^K{}_J
- \delta_\phi {\phi}_{JK} \wedge  \ovr{A}^K{}_{I})
\wedge \ovr{\star (e^I\wedge e^J)}
\nppp&&
+
2\delta\phi\,\phi\ovr{F}_{IJ}\wedge  \star (e^I\wedge e^J)
\nppp
&\deq&
\delta_\phi {\phi}_{IJ}
\wedge
\ovr{\DDD\star (e^I\wedge e^J)}
+
2\delta\phi\,\phi^{-1}\ovr{F}_{IJ}\wedge\ovr{\star (e^I\wedge e^J)}.
\label{dReephi}
\end{eqnarray}

Using \eqref{torfr}  and \eqref{dReephi} implied from the above $A_{IJ}$ variations,
the $\phi$-variation equation
\begin{eqnarray}
\delta_\phi[\ovr{F_{IJ} \wedge \star (e^I\wedge e^J)}]
=0
\label{dReephieq}
\end{eqnarray}
then yields
\begin{eqnarray}
\ovr{F_{IJ} \wedge \star (e^I\wedge e^J)}
=0
\label{dReephieqn}
\end{eqnarray}
which is equivalent to $\ovr{R}=0$ and is satisfied if $\ovr{g}$ satisfies the vacuum Einstein equation, as will be established in \eqref{eineq} below.

From \eqref{be}, \eqref{dARab}, \eqref{dpsiab}, \eqref{Reeh} we have the variation with respect to $\theta$ as follows
\begin{eqnarray}
\delta_\theta[\und{F}_{IJ} \wedge \und{ (e^I\wedge e^J)}]
&=&
(\dd \delta_\theta {\theta}_{IJ}
+ \delta_\theta {\theta}_{IK} \wedge \und{A}^K{}_J
- \delta_\theta {\theta}_{JK} \wedge  \und{A}^K{}_{I})
\wedge \und{ (e^I\wedge e^J)}
\nppp&&
+
\delta(\theta^2)\,\und{F}_{IJ}\wedge (e^I\wedge e^J)
\nppp
&\deq&
\delta_\theta {\theta}_{IJ}
\wedge
\und{\DDD (e^I\wedge e^J)}
+
2\delta\theta\,\theta^{-1}\und{F}_{IJ}\wedge\und{ (e^I\wedge e^J)}.
\label{dReephi2}
\end{eqnarray}

Using \eqref{dss2} and \eqref{dReephi2} implied from the above $A_{IJ}$ variations,
the $\theta$-variation equation \begin{eqnarray}
\delta_\theta[\und{F}_{IJ} \wedge \und{ (e^I\wedge e^J)}]
=0
\label{dReephieq2}
\end{eqnarray}
then yields
\begin{eqnarray}
\und{F}_{IJ} \wedge \und{ (e^I\wedge e^J)}
=0
\label{dReephieqn2}
\end{eqnarray}
which is also satisfied by \eqref{dss2} through \eqref{DTI} with zero torsion.

\subsection{Variations with respect to the tetrad $e^I$}

From  \eqref{Dsee} and \eqref{dARab} we have
\begin{eqnarray}
\delta_e\ovr{F_{IJ} \wedge \star (e^I\wedge e^J)}
&=&
(\dd \delta_e \phi_{IJ}
+ \delta_e \phi_{IK} \wedge \ovr{A}^K{}_J
- \delta_e \phi_{JK} \wedge  \ovr{A}^K{}_{I})
\wedge \ovr{\star (e^I\wedge e^J)}
\nppp
&\deq&
\delta_e \phi_{IJ}
\wedge
\phi^2{\DDD\star (e^I\wedge e^J)}.
\label{dARabe}
\end{eqnarray}

When the torsion-free relation \eqref{torfr} as a result of $A$-variation is applied, the above variation \eqref{dARabe} yields
\begin{eqnarray}
\delta_e\ovr{F_{IJ} \wedge \star (e^I\wedge e^J)}
&=&
0
\label{dARabe0}
\end{eqnarray}

From \eqref{delsee} and \eqref{delsee2} we have
\begin{eqnarray}
\delta_e\,\ovr{\star (e^I\wedge e^J)}
&=&
\delta e_K \wedge\phi\, \ovr{\star (e^I\wedge e^J\wedge e^K)}.
\label{delesee}
\end{eqnarray}

It follows from \eqref{dARabe0} and \eqref{delesee} that
\begin{eqnarray}
\delta_e[\ovr{F_{IJ} \wedge \star (e^I\wedge e^J)}]
&\deq&
\delta_e\,\ovr{\star (e^I\wedge e^J)} \wedge   \ovr{F}_{IJ}
\nppp
&\deq&
\delta e_K \wedge\phi\, \ovr{\star (e^I\wedge e^J\wedge e^K)}
\wedge   \ovr{F}_{IJ}
\nppp
&\deq&
\delta e^K \wedge\phi\,  \ovr{G}_K
\label{dReeee2}
\end{eqnarray}
in terms of the Einstein 3-forms
\begin{eqnarray}
G_K
&=&
{F}_{IJ}\wedge \ii_K{\star (e^I\wedge e^J)}.
\label{GK}
\end{eqnarray}

Therefore the $e$-variational equation for the first term in \eqref{S} yields the the vacuum Einstein equation
\begin{eqnarray}
\ovr{G}_K = 0
\label{eineq}
\end{eqnarray}

From \eqref{dARab} and \eqref{Dseee} we have
\begin{eqnarray}
\delta_e\und{F}_{IJ} \wedge \und{ (e^I\wedge e^J)}
&=&
(\dd \delta_e \theta_{IJ}
+ \delta_e \theta_{IK} \wedge \und{A}^K{}_J
- \delta_e \theta_{JK} \wedge  \und{A}^K{}_{I})
\wedge \und{ (e^I\wedge e^J)}
\nppp
&\deq&
\delta_e \theta_{IJ}
\wedge
\theta^2{\DDD (e^I\wedge e^J)}.
\label{dARabeh}
\end{eqnarray}

When the torsion-free relation \eqref{dss2} as a result of $A$-variation is applied, the above variation \eqref{dARabeh} yields
\begin{eqnarray}
\delta_e\und{F_{IJ} \wedge (e^I\wedge e^J)}
&\deq&
0.
\label{dARabeh0}
\end{eqnarray}

It follows that
\begin{eqnarray}
\delta_e[\und{F_{IJ} \wedge (e^I\wedge e^J)}]
&\deq&
\delta_e\,\und{(e^I\wedge e^J)} \wedge   \und{F}_{IJ}
\nppp
&\deq&
\delta\,e^I\wedge 2\theta\,\und{e}^J \wedge   \und{F}_{IJ}
\nppp
&\deq&
0
\label{GKh}
\end{eqnarray}
using \eqref{DTI} satisfied by the torsion-free connection of $A_{IJ}$, and hence $\und{A}_{IJ}$.

\subsection{Expansions in coordinates}

In spacetime coordinates $(x^\alpha)$, the tetrad co-frame and frame are expanded as
\begin{eqnarray}
e^I
&=&
e_\alpha{}^I \dd x^\alpha
\label{cofr1}
\ppp
\tilde{e}_I
&=&
e^\alpha{}_I \partial_\alpha
\label{cofr2}
\end{eqnarray}
respectively, so that
\begin{eqnarray*}
\ii_{I} \dd x^\alpha
&=&
\ii_{I} e^\alpha{}_J e^J
=
e^\alpha{}_I
\end{eqnarray*}
and
\begin{eqnarray}
\dd x^\alpha
&=&
e^\alpha{}_I e^I
\label{coordfr1}
\ppp
\frac{\partial}{\partial x^\alpha}
&=&
e_\alpha{}^I \tilde{e}_I.
\label{coordfr2}
\end{eqnarray}

The spacetime metric \eqref{gab} then reads
\begin{eqnarray}
g_{\alpha\beta}
=
e_\alpha{}_I e_\beta{}^I
\label{cgab}
\end{eqnarray}
using the tetrad (components) $ e_\alpha{}^I $ with inverse $ e^\alpha{}_I $ so that \eqref{ee} becomes
\begin{eqnarray}
e_\alpha{}^I e^\alpha{}_J
=
\delta^I_J.
\label{cee}
\end{eqnarray}

From \eqref{cgab} we also have
\begin{eqnarray}
e^\alpha{}_I e_\beta{}^I
=
\delta^\alpha_\beta
\label{ceee}
\end{eqnarray}

The contravariant metric \eqref{cginvab} then reads
\begin{eqnarray}
g^{\alpha\beta}
=
e^\alpha{}_I e^\beta{}^I.
\label{cginvab}
\end{eqnarray}

We adopt the LC antisymmetric symbols with tetrad indices $\epsilon_{IJKL}$  and coordinate indices $\epsilon_{\alpha\beta\gamma\delta}$ subject to
$\epsilon_{0123}=1$.
It follows that
\begin{eqnarray*}
\epsilon
&=&
\det(e_\alpha{}^I)
=
-\frac{1}{4!}\,
\epsilon^{\alpha\beta\gamma\delta}\,
\epsilon_{IJKL}
e_\alpha{}^I e_\beta{}^J e_\gamma{}^K e_\delta{}^L
\end{eqnarray*}
and
\begin{eqnarray*}
\det(e_\alpha{}_I)
&=&
\det(e_\alpha{}^J) \det(\eta_{IJ})
=
-\epsilon.
\end{eqnarray*}

Therefore, we also have
\begin{eqnarray*}
g
&=&
\frac{1}{4!}\,
\epsilon^{\alpha\beta\gamma\delta}\,
e_\alpha{}_I e_\beta{}_J e_\gamma{}_K e_\delta{}_L
\times
\epsilon^{\alpha'\beta'\gamma'\delta'}\,
e_{\alpha'}{}^I e_{\beta'}{}^J e_{\gamma'}{}^K e_{\delta'}{}^L
\nppp
&=&
\det(e_\alpha{}_I) \times \det(e_\alpha{}^I)
=
-\epsilon^2
\end{eqnarray*}
and hence
\begin{eqnarray}
\epsilon
&=&
\sqrt{-g}.
\end{eqnarray}

The spacetime volume 4-form is given by
\begin{eqnarray}
\star 1
&=&
\frac{1}{4!}\,\epsilon_{IJKL}
e^I \wedge e^J \wedge e^K \wedge e^L
\nppp
&=&
\frac{1}{4!}\,\epsilon_{IJKL}
e_\alpha{}^I e_\beta{}^J e_\gamma{}^K e_\delta{}^L\,
\dd x^\alpha \wedge \dd x^\beta \wedge \dd x^\gamma \wedge \dd x^\delta.
\label{star1a}
\end{eqnarray}

On the other hand, we have
\begin{eqnarray}
\star 1
&=&
\sqrt{-g}\,
\dd x^0 \wedge \dd x^1 \wedge \dd x^2 \wedge \dd x^3
\nppp
&=&
\frac{\epsilon}{4!}\,\epsilon_{\alpha\beta\gamma\delta}\,
\dd x^\alpha \wedge \dd x^\beta \wedge \dd x^\gamma \wedge \dd x^\delta.
\label{star1b}
\end{eqnarray}

Note that in the above $\epsilon\,\epsilon_{\alpha\beta\gamma\delta}$ is a weight 0 antisymmetric tensor.

Comparing \eqref{star1a} and \eqref{star1b} we see that
\begin{eqnarray}
\epsilon\,\epsilon_{\alpha\beta\gamma\delta}\,
&=&
\epsilon_{IJKL}
e_\alpha{}^I e_\beta{}^J e_\gamma{}^K e_\delta{}^L.
\label{esp}
\end{eqnarray}

The components  of $A_{IJ}$ are given by
\begin{eqnarray}
A_{IJ}
&=&
A_{\alpha IJ}\dd x^\alpha
\label{Acmp}
\end{eqnarray}
and likewise for $\Gamma_{IJ}$, $\phi_{IJ}$ and so forth.

Then, from \eqref{Rab} and \eqref{Acmp} we have
\begin{eqnarray*}
F_{IJ}
&=&
\frac12\,
F_{\alpha\beta\,IJ}\,
\dd x^\alpha \wedge \dd x^\beta
\end{eqnarray*}
where
\begin{eqnarray}
F_{\alpha\beta\,IJ}
&=&
\partial_{\alpha} A_{\beta IJ}
-
\partial_{\beta} A_{\alpha IJ}
+
A_{\alpha\, IK} A_{\beta}{}^K{}_J
-
A_{\beta\, IK} A_{\alpha}{}^K{}_J.
\label{FabIJ}
\end{eqnarray}

By expanding \eqref{wab} with the relation
\begin{eqnarray*}
\dd e_I
&=&
-e_{\alpha I, \beta}\, e^\alpha{}_J e^\beta{}_K \, e^J \wedge e^K
\end{eqnarray*}
we obtain
\begin{eqnarray}
\Gamma_{\mu\,IJ}
&=&
\frac12\,\big[
e_\mu{}^K
e_{\alpha K, \beta}\,
(e^\alpha{}_I e^\beta{}_J
-
e^\alpha{}_J e^\beta{}_I )
+
(
e_{\mu I, \alpha}
-
e_{\alpha I, \mu}
)\,
e^\alpha{}_J
-
(
e_{\mu J, \alpha}
-
e_{\alpha J, \mu}
)\,
e^\alpha{}_I
\big]
\label{wab3}
\end{eqnarray}

From \eqref{psiab}, \eqref{cofr1} and \eqref{cofr2} we find that
\begin{eqnarray}
\phi_{\alpha IJ}
&=&
\phi^{-1}
\phi_{,\beta}\, (
e^\beta{}_J \,e_\alpha{}_I
-
e^\beta{}_I \, e_\alpha{}_J ).
\label{PhiaIJ}
\end{eqnarray}
%


\subsection{Time gauge decompositions}

In canonical gravity, the ADM decomposition of spacetime metric into the lapse function $N$, shift vector $N^a$, and spatial metric
\begin{eqnarray}
h_{ab}
&=&
g_{ab} = e_a{}^i e_b{}^i
\label{h2e}
\end{eqnarray}
corresponds the ``time gauge'' in which the tetrad 1-forms  $e_\alpha{}^I$ are expressed in terms of the lapse, shift, and the triad 1-forms $e_a{}^i$, as  the spatial part of the tetrad, with
\begin{subequations}
\begin{eqnarray}
e_\alpha{}^0
&=&
N\, \delta_\alpha^t
\label{e0}
\ppp
e_t{}^i
&=&
N^a e_a{}^i.
\label{e0i}
\end{eqnarray}
\label{tetr1}
\bppp
\end{subequations}
Accordingly, the tetrad vectors $\tilde{e}^\alpha{}_I$ are given expressed in terms of the lapse, shift, and the triad vectors $e^a{}_i$, as  the spatial part of the tetrad, with
\begin{subequations}
\begin{eqnarray}
e^t{}_I
&=&
\frac{1}{N}\,\delta^0_I
\label{ee0}
\ppp
e^a{}_0
&=&
-\frac{N^a}{N}
\label{ee0i}
\end{eqnarray}
\label{tetr2}
\bppp
\end{subequations}
with their spatial part giving rise to the contravariant spatial metric
$h^{ab} = e^a{}_i e^b{}_i$ so that $h^{ac} \, h_{cb} = \delta^a_b$.

In this gauge, \eqref{ee0} means $\tilde{e}_0$ is a unit vector normal to the equal-time ($t$) hypersurface and \eqref{ee0i} means unit vectors $\tilde{e}_i$ are perpendicular to $\tilde{e}_0$ and span the tangent space of the equal-time hypersurface.

The spatial LC antisymmetric symbols are given by
$\epsilon_{i j k} := \epsilon_{0 I J K}$
and
$\epsilon_{a b c} := \epsilon_{t a b c}$
so that $\epsilon_{123}=1$.

From \eqref{esp} and \eqref{tetr1} we have
\begin{eqnarray}
e\, \epsilon_{a b c}\,
&=&
\epsilon_{i j k}
e_a{}^i e_b{}^j e_c{}^k
\label{eN}
\end{eqnarray}
where
\begin{eqnarray}
e
&:=&
\det(e_a{}^i)
=
\sqrt{h}
=
\frac{\epsilon}{N}
\label{dettriad}
\end{eqnarray}
is the determinant of the triad.

The densitized triad is given by
\begin{eqnarray*}
E^a{}_i
&=&
e\,  e^a{}_i
\end{eqnarray*}
satisfying
\begin{eqnarray*}
E
:=
\det(E^a{}_i)
=
e^2
=
h.
\end{eqnarray*}
As per standard notation, the inverse of $E^a{}_i$ is denoted by the weight $-1$ $E_a{}^i$.

In terms of the LC connection \eqref{wab},
the LC spin connection 1-forms are given by
\begin{eqnarray}
\Gamma_{a i}
&=&
-
\frac{1}{2}\,
\epsilon_{ijk}
\Gamma_{a jk}
\label{Gai}
\end{eqnarray}
used to define the LC spin covariant derivative $\nabla$ compatible with the triad such that
\begin{subequations}
\begin{eqnarray}
\nabla_a e_b{}^i
&=&
0
=
\nabla_a e^b{}_i
\ppp
\nabla_a E_b{}^i
&=&
0
=
\nabla_a E^b{}_i.
\end{eqnarray}
\label{cmpbly}
\bppp
\end{subequations}


Using \eqref{PhiaIJ}, \eqref{tetr1} and \eqref{tetr2} we find
\begin{subequations}
\begin{eqnarray}
\phi_{a i 0}
&=&
\frac{\sqrt{E}}{N}\,
\phi^{-1}
(\dot\phi-N^c\phi_{,c})\,E_{a i}
\label{ppai0}
\ppp
\phi_{a i j}
&=&
\phi^{-1}
\phi_{,b}\, (
E_a{}_i E^b{}_j
-
E_a{}_j E^b{}_i
)
\label{ppaij}
\ppp
\phi_{t i 0}
&=&
\sqrt{E}\,
\phi^{-1}
\Big(
\frac{N^a}{N}\,
\dot{\phi}
-
\frac{N^a N^b}{N}\,
\phi_{,b}
+
N\,\phi^{,a}
\Big)
E_a{}_i
\label{ppti0}
\ppp
\phi_{t i j}
&=&
\phi^{-1}
\phi_{,b}\, N^a (
E_a{}_i E^b{}_j
-
E_a{}_j E^b{}_i
)
\label{pptij}
\end{eqnarray}
\bppp
\end{subequations}

\subsection{Canonical analysis of the conformal Holst action}

Without the loss of generality, we can always perform a conformal transformation also that $\theta$ becomes
\begin{eqnarray}
\theta
=
\frac{1}{\sqrt{\beta}}
\label{cfx2}
\end{eqnarray}
for some constant $\beta$.
Although this amounts to a partial conformal gauge fixing, the primary conformal scalar field $\phi$ remains completely free. As will become clear below, this has an important effect of making $\beta$ not a fixed constant since under a certain constant conformal transformation the value of $\beta$ will transform too.

Using \eqref{FP}, \eqref{FQ}, and \eqref{cfx2}, the action \eqref{S} becomes
\begin{eqnarray}
S
&=&
\frac1{2}\int
\big[
\phi^2
\ovr{F}_{IJ} \wedge \star (e^I\wedge e^J)
-
\frac{1}{\beta}\,F_{IJ} \wedge e^I \wedge e^J
\big].
\label{Sha}
\end{eqnarray}

Using
\begin{eqnarray*}
\dd^4x
=
\dd x^0 \wedge \dd x^1 \wedge \dd x^2 \wedge \dd x^3
\end{eqnarray*}
we have the expansions
\begin{gather*}
{F_{IJ} \wedge {\star (e^I\wedge e^J)}}
=
{\epsilon\, e^\alpha{}_I  e^\beta{}_J
F_{\alpha\beta}{}^{IJ}}\,\dd^4x
\ppp
{F_{IJ}\wedge e^I\wedge e^J}
=
-\frac12\,\epsilon^{IJ}{}_{KL}
{\epsilon\,e^\alpha{}_{I} e^\beta{}_{J}
F_{\alpha\beta}{}^{KL}}\,\dd^4x
\end{gather*}
so that \eqref{Sha} can be written as
$S=\int\LL\,\dd^4x$
in terms of the Lagrangian density
\begin{eqnarray}
\LL
&=&
\frac{1}{2}\,
\epsilon\, e^\alpha{}^I  e^\beta{}^J
\Big[
\phi^2
\ovr{F}_{\alpha\beta}{}_{IJ}
+
\frac{1}{2\beta}\,
\epsilon_{IJ}{}^{KL}F_{\alpha\beta}{}_{KL}
\Big]
\nppp
&=& 
E^a{}^i
\HH_{t a}{}_{i 0}
-
N^a E^b{}^i
\HH_{ab}{}_{i 0}
+
\frac{N}{2e}\,E^a{}^i  E^b{}^j
\HH_{ab}{}_{i j}
\label{Lha}
\end{eqnarray}
where
\begin{eqnarray}
\HH_{\alpha\beta}{}_{IJ}
&=&
\phi^2 \ovr{F}_{\alpha\beta}{}_{IJ}
+
\frac{1}{2\beta}\,
\epsilon_{IJ}{}^{KL}F_{\alpha\beta}{}_{KL}.
\label{fff}
\end{eqnarray}

The first term of \eqref{Lha} can be expanded to be
\begin{eqnarray}
E^{a i} \HH_{t a\,i 0}
&=&
E^{a i} \partial_{t} \big(
\phi^2 \ovr{A}_{a\, i 0}
-
\frac{1}{2\beta}\,
\epsilon_{ijk} A_{a\, jk}
\big)
-
2\,E^{a i} \ovr{A}_{a\, i 0}\,\phi\,\partial_{t}\phi
+
E^{a i} \GG_{a i}
\label{Efff1}
\end{eqnarray}
where
\begin{eqnarray}
\GG_{a i}
&=&
-
\partial_{a} \big(
\phi^2 \ovr{A}_{t\, i 0}
-
\frac{1}{2\beta}\,
\epsilon_{ijk} A_{t\, jk}
\big)
+
2 \ovr{A}_{t\, i 0}\,\phi\,\phi_{,a}
\nppp
&&
+
\phi^2
\big(
\ovr{A}_{t\, i m} \ovr{A}_{a\, m 0}
-
\ovr{A}_{a\, i m} \ovr{A}_{t\, m 0}
\big)
-
\frac{1}{\beta}\,
\epsilon_{ijk}
(
A_{t\, j 0} A_{a\, k 0}
-
A_{t\, j m} A_{a\, k m}
).
\label{Efff2}
\end{eqnarray}
%


\subsubsection{Canonical conformal connection variables}

From \eqref{Efff1}, we see that $E^{a i} \HH_{t a\,i 0}$ contains the time derivative of the quantity
\begin{eqnarray}
A_{a i}
&:=&
\beta\phi^2 \ovr{A}_{a i 0}
-
\frac{1}{2}\,
\epsilon_{ijk}
A_{a jk}
\nppp
&=&
\beta\phi^2 A_{a i 0}
+
\beta\phi^2 \phi_{a i 0}
-
\frac{1}{2}\,
\epsilon_{ijk}
A_{a jk}
\label{Aai}
\end{eqnarray}
where \eqref{ppai0} has been used, that allows $A_{a i}$ to be identified as the canonical connection variable with ${E^{a i}}/{\beta}$
as the conjugate momentum.

As a result of the variational equations \eqref{dss12}, leading to the torsion free condition \eqref{torfr}, we will eliminate torsion by setting
\begin{eqnarray}
A_{\alpha IJ}
&=&
\Gamma_{\alpha IJ}
\label{A2Gam}
\end{eqnarray}
with the LC connection \eqref{wab3} in \eqref{fff}.
Consequently, the dynamical structure becomes closer to that of the ADM second-order description, where the time evolution of the spatial metric \eqref{h2e} is related to the extrinsic curvature tensor given by
\begin{eqnarray}
K_{ab}
=
\frac1{2N}
\big(- \dot{h}_{ab} + N_{a;b} + N_{b;a}\big)
\label{defKij}
\end{eqnarray}
with a trace denoted by $K=h^{ab} K_{ab}$.

Analogously, here the time evolution of the triad is related to the extrinsic curvature 1-forms given by
\begin{eqnarray}
K_{a i}
&=&
-K_{ab}e^b{}_i.
\label{Ke}
\end{eqnarray}

Conversely, the extrinsic curvature tensor can be recovered with
$
K_{ab}
=
-K_{(a}{}^i e_{b)}{}^i
$
or simply
$
K_{ab}
=
-K_{a}{}^i e_{b}{}^i
$
subject to the Gauss constraint. See \eqref{gausscons} below.

Furthermore, it is useful to introduce the {\it conformal extrinsic curvature} 1-forms
\begin{eqnarray}
C_{a i}
&:=&
\phi^2
(K_{a i} + \phi_{a i 0})
\label{Kiax}
\end{eqnarray}
as they will play an important dynamical role.

From the LC connection in spatial coordinates
\begin{equation*}
\Gamma_{c a b}
=
\frac12
\big(
e_c{}^k e_{a k,b}
+
e_a{}^k e_{c k,b}
+
e_c{}^k e_{b k,a}
+
e_b{}^k e_{c k,a}
-
e_a{}^k e_{b k,c}
-
e_b{}^k e_{a k,c}
\big)
\end{equation*}
we can evaluate
\begin{eqnarray*}
(N_{a;b}+N_{b;a})e^{b}{}_i
&=&
N^b{}_{,a} e_{b i}
+
N^b e_{a i,b}
+
N^b{}_{,c} e^{c}{}_i e_{a j} e_{b j}
-
N^b e_{a j} e_{c j} e^{c}{}_{i,b}
\end{eqnarray*}

Substituting \eqref{defKij} and the above relation into \eqref{Ke}, one finds that the expression
\begin{eqnarray}
\Gamma_{a\,i0}
&=&
\frac{1}{2N}\,
\big[
e_{a i, t}
+
e_a{}^j e^b{}_i e_{b j, t}
-
N^b{}_{, a} e_{b i}
-
N^b e_{a i, b}
-
N^b{}_{, c} e^c{}_i e_{a j} e_{b j}
+
N^b e^c{}_{i, b} e_{a j} e_{c j}
\big]
\label{Gai0}
\end{eqnarray}
obtained from \eqref{wab3}, \eqref{tetr1}, and \eqref{tetr2}, to reduce to the identity
\begin{eqnarray}
\Gamma_{a\,i0}
&=&
K_{ai}
\label{excurv}
\end{eqnarray}

Therefore, using \eqref{ppai0}, \eqref{Gai}, \eqref{A2Gam}, and \eqref{excurv}, Eq. \eqref{Aai} yields the {\it conformal spin connection} variable
\begin{eqnarray}
A_{a i}
&=&
\Gamma_{a i} + \beta C_{a i}.
\label{AAai}
\end{eqnarray}

We further introduce the curvature 2-forms
\begin{eqnarray}
F_{ab\,i}
&=&
2\, \partial_{[a} A_{b] i} + \epsilon_{ijk} A_{a j} A_{b k}
\label{Fkab}
\end{eqnarray}
of the conformal spin connection \eqref{AAai}.

\subsubsection{Canonical conformal scalar variables}

Using \eqref{bwab}, \eqref{ppai0}, \eqref{Lha}, \eqref{fff},
and \eqref{A2Gam}, let us now consider the time derivative of the scalar $\phi$ in
$E^a{}^i \HH_{t a}{}_{i 0}$ as follows,
\begin{eqnarray}
-2\,E^a{}^i \ovr{A}_{a\, i 0}\,\phi\,\dot\phi
&=&
-2\,E^a{}^i
(A_{a\, i 0} + \phi_{a\, i 0})\,
\phi\,\dot\phi
\nppp
&=&
\big[
-2\,\phi\,
E^a{}^i
K_{a\, i}
-
\frac{6\,\sqrt{E}}{N}\,(\dot\phi-N^c\phi_{,c})
\big]
\dot\phi
\nppp
&=&
\pi_\phi\,\dot\phi
\end{eqnarray}
where we have identified
\begin{eqnarray}
\pi_\phi
&=&
-2\,E^a{}^i \ovr{A}_{a\, i 0}\,\phi
\nppp
&=&
2\,\sqrt{E}\phi\,K
-
\frac{6\,\sqrt{E}}{N}\,(\dot\phi-N^c\phi_{,c})
\label{piphi}
\end{eqnarray}
as the canonical momentum of $\phi$.

\subsubsection{Conformal constraint}


It follows from \eqref{Kiax} and \eqref{piphi} that
\begin{eqnarray}
C_{a i} E^{a i}
&=&
\phi^2 K_{a i} E^{a i}
+
\frac{\sqrt{E}}{N}\,\phi\,(\dot\phi-N^c\phi_{,c})\,E_{a i} E^{a i}
\nppp
&=&
-
\sqrt{E}\,\phi^2 K
+
\frac{3\,\sqrt{E}\phi}{N}\,(\dot\phi-N^c\phi_{,c})
\nppp
&=&
-\frac{1}{2}\,\phi\,\pi_\phi.
\label{confconstr0}
\end{eqnarray}

This gives rise to the conformal constraint
\begin{eqnarray}
\CC
&=&
\phi\,\pi_\phi
+
2\,C_{a i} E^{a i}
\label{confcons}
\end{eqnarray}

\subsubsection{Gauss constraint}

By using \eqref{bwab}, \eqref{cmpbly}, \eqref{A2Gam}, and dropping a total divergence arising from the first term of \eqref{Efff2}, this equation yields
\begin{eqnarray}
E^{a i} \GG_{a i}
&=& 
2 E^{a i} \Gamma_{t\, i 0}\,\phi\,\phi_{,a}
+
2 E^{a i} \phi_{t\, i 0}\,\phi\,\phi_{,a}
+
E^{a i} \phi^2
K_{a j}
\Gamma_{t\, i j}
+
E^{a i} \phi^2
\phi_{a\, j 0}
\Gamma_{t\, i j}
\nppp
&&
+
E^{a i} \phi^2
\big(
\phi_{t\, i j} K_{a j}
-
\phi_{a\, i j} \Gamma_{t\, j 0}
\big)
+
E^{a i} \phi^2
\big(
\phi_{t\, i j} \phi_{a\, j 0}
-
\phi_{a\, i j} \phi_{t\, j 0}
\big)
\nppp
&&
-
\frac{1}{2\beta}\,
\epsilon_{mjk}
\Gamma_{a i m}
\big(
\Gamma_{t\, j k}
E^{a}{}_{i}
+
\Gamma_{t\, k i}
E^{a}{}_{j}
+
\Gamma_{t\, i j}
E^{a}{}_{k}
\big).
\label{Efff3}
\end{eqnarray}

Upon inspecting the symmetry of
$(
\Gamma_{t\, j k}
E^{a}{}_{i}
+
\Gamma_{t\, k i}
E^{a}{}_{j}
+
\Gamma_{t\, i j}
E^{a}{}_{k}
)$
in the last term of \eqref{Efff3}
with respect to $(i, j, k)$, we see that it is proportional to $\epsilon_{ijk}$ and hence this term vanishes as a result of
\begin{eqnarray*}
\frac{1}{2}\,
\epsilon_{mjk}
\epsilon_{ijk}
\Gamma_{a i m}
&=& 
\delta_{i m}
\Gamma_{a i m}
=
0
\end{eqnarray*}
and so Eq. \eqref{Efff3} becomes
\begin{eqnarray}
E^{a i} \GG_{a i}
&=& 
2 E^{a}{}_{i}\,\phi\,\phi_{,a} \ovr{\Gamma}_{t\, i 0}
-
E^{a}{}_{i} \phi^2
\phi_{a\, i j} \ovr{\Gamma}_{t\, j 0}
+
E^{a}{}_{i}
\ovr{K}_{a j}
\ovr{\Gamma}_{t\, i j}
+
\frac{1}{\beta}\,
\epsilon_{ijk}
E^{a}{}_{i}
K_{a j}
\Gamma_{t\, k 0}
\label{Efff4}
\end{eqnarray}
where \eqref{Kiax} has been used.

From \eqref{ppaij} we have
\begin{eqnarray*}
E^{a i} \phi^2 \phi_{a i j}
&=&
\phi\,
\phi_{,c}\, (
E^c{}_j \, E_a{}_i E^{a i}
-
E^c{}_i \, E_a{}_j E^{a i})
\nppp
&=&
2\phi\,\phi_{,c}\,E^c{}_j.
\end{eqnarray*}

Substituting this into \eqref{Efff4} we get
\begin{eqnarray}
E^{a i} \GG_{a i}
&=& 
E^{a i}
\ovr{K}_{a j}
\ovr{\Gamma}_{t\, i j}
+
\frac{1}{\beta}\,
\epsilon_{ijk}
E^{a}{}_{i}
K_{a j}
\Gamma_{t\, k 0}.
\label{Efff5}
\end{eqnarray}

From \eqref{Kiax} and \eqref{ppai0}, we have
\begin{eqnarray*}
\epsilon_{ijk}
E^{a}{}_{i} K_{a j}
&=&
\phi^{-2}
\epsilon_{ijk}
E^{a}{}_{i} C_{a j}
-
\epsilon_{ijk}
E^{a}{}_{i} \phi_{a j 0}
\nppp
&=&
\phi^{-2}
\epsilon_{ijk}
E^{a}{}_{i} C_{a j}
-
\frac{\sqrt{E}}{N\,\phi}\,(\dot\phi-N^c\phi_{,c})\,
\epsilon_{ijk}
E^{a}{}_{i}
E_{a j}
\nppp
&=&
\phi^{-2}
\epsilon_{ijk}
E^{a}{}_{i} C_{a j}
\end{eqnarray*}

Substituting this into \eqref{Efff5}, we arrive at
\begin{eqnarray}
E^{a i} \GG_{a i}
&=& 
E^{a}{}_{i}
\ovr{K}_{a j}
(
\Gamma_{t\, i j}
+
\phi_{t\, i j}
)
+
\frac{1}{\beta}\,
\epsilon_{ijk}
E^{a}{}_{i}
\ovr{K}_{a j}
\Gamma_{t\, k 0}
\nppp
&=& 
\frac{1}{2}\,
\epsilon_{ijk}
\epsilon_{lmk}
E^{a}{}_{i}
\ovr{K}_{a j}
(
\Gamma_{t\, l m}
+
\phi_{t\, l m}
)
+
\frac{1}{\beta}\,
\epsilon_{ijk}
E^{a}{}_{i}
\ovr{K}_{a j}
\Gamma_{t\, k 0}
\nppp
&=& 
\epsilon_{ijk}
E^{a}{}_{i}
\ovr{K}_{a j}\,
\big[
\frac{1}{2}\,
\epsilon_{klm}
(
\Gamma_{t\, l m}
+
\phi_{t\, l m}
)
+
\frac{1}{\beta}\,
\Gamma_{t\, k 0}
\big]
\label{Efff6}
\end{eqnarray}

Since $\Gamma_{t\, I J}$ are non-dynamical variables without conjugate momenta, they behave like Lagrangian multipliers in \eqref{Efff6}, giving rise to the Gauss constraint
\begin{eqnarray}
\GG_i
&=&
\epsilon_{ijk}C_{a j}E^a{}_{k}
=
\epsilon_{ijk} C_{a j} E^a{}_{k}
\nppp
&=&
\beta^{-1}
(
E^a{}_{i,a} + \epsilon_{ijk}A_{a j}E^a{}_{k}
)
\nppp
&=&
\beta^{-1}D_a E^a{}_{i}
\label{gausscons}
\end{eqnarray}
in terms of the spin covariant derivative $D_a$ with respect to the (non-LC) spin connection $A_a{}^i$. It can be noted that in the limiting case with
$\beta\to0$
then
$D\to\nabla$
and
$A_a{}^i\to\Gamma_a{}^i$, the LC spin connection \eqref{Gai}.

\subsubsection{Diffeomorphism constraint}

By substituting \eqref{AAai} into \eqref{Fkab}, we have
\begin{eqnarray}
F_{ab\,i}
&=&
R_{ab\,i}
+
2\beta\,{\nabla}^{}_{[a} C_{b] i}
+
\beta^2 \epsilon_{ijk} C_{a j} C_{b k}
\label{Fkab2}
\end{eqnarray}
where
\begin{eqnarray}
R_{ab\,i}
&=&
2\, \partial_{[a} \Gamma_{b] i} + \epsilon_{ijk} \Gamma_{a j} \Gamma_{b k}
\label{Rkab}
\end{eqnarray}
is the curvature 2-forms of the LC spin connection $\Gamma_a{}^i$.
Since it is torsion-free, the first Bianchi identity \eqref{DTI} gives rise to the identiy
\begin{eqnarray}
E^{b i} R_{ab\,i}
&=&
0
\label{ER0}
\end{eqnarray}
which in turn allows Eq. \eqref{Fkab2} to yield the relation
\begin{eqnarray}
E^{b i} F_{ab\,i}
&=&
2\beta\,E^{b i} {\nabla}^{}_{[a} C_{b] i}
+
\beta^2 E^{b i} \epsilon_{ijk} C_{a j} C_{b k}.
\label{EFkab2}
\end{eqnarray}

Furthermore, it follows from \eqref{gausscons} and \eqref{Fkab2} that
\begin{eqnarray}
\epsilon_{i j k}
F_{ab\,k}
E^a{}^i  E^b{}^j
&=&
\epsilon_{i j k}
R_{ab\,k}
E^a{}^i  E^b{}^j
+
2\beta^2
C_{[a}{}^{i} C_{b]}{}^{j}
E^a{}^i  E^b{}^j
-
2\,\beta\, E^a{}^i
{\nabla}^{}_{a}\, \GG_i.
\label{epsiFPP1x}
\end{eqnarray}

On applying \eqref{A2Gam}, the second  term in Eq. \eqref{fff} becomes
\begin{eqnarray*}
\HH_{a b\,i 0}
&=& 
\partial_{a} (\phi^2 \ovr{A}_{b\, i 0})
-
\partial_{b} (\phi^2 \ovr{A}_{a\, i 0})
-
2\,\ovr{A}_{b\, i 0}\,\phi\,\partial_{a}\phi
+
2\,\ovr{A}_{a\, i 0}\,\phi\,\partial_{b}\phi
\nppp
&&
+
\phi^2 \ovr{A}_{a\, i j} \ovr{A}_{b}{}^j{}_0
-
\phi^2 \ovr{A}_{b\, i j} \ovr{A}_{a}{}^j{}_0
-
\frac{1}{2\beta}\,
\epsilon_{ijk}
(
A_{a\, j 0} A_{b}{}^0{}_k
-
A_{b\, j 0} A_{a}{}^0{}_k
)
-
\frac{1}{2\beta}\,
R_{a b\,i}.
\end{eqnarray*}

Using \eqref{piphi} and \eqref{ER0}, the above yields
\begin{eqnarray*}
E^b{}^i
\HH_{a b\,i 0}
&=& 
E^b{}^i
\partial_{a} (\phi^2 \ovr{A}_{b\, i 0})
-
E^b{}^i
\partial_{b} (\phi^2 \ovr{A}_{a\, i 0})
+
\pi_\phi \partial_{a}\phi
+
2\,E^b{}^i
\ovr{A}_{a\, i 0}\,\phi\,\partial_{b}\phi
\nppp
&&
+
\phi^2 E^b{}^i
\ovr{A}_{a\, i j} \ovr{A}_{b}{}^j{}_0
-
\phi^2 E^b{}^i
\ovr{A}_{b\, i j} \ovr{A}_{a}{}^j{}_0
-
\frac{1}{2\beta}\,
E^b{}^i
\epsilon_{ijk}
(
A_{a\, j 0} A_{b}{}^0{}_k
-
A_{b\, j 0} A_{a}{}^0{}_k.
)
\end{eqnarray*}
By using \eqref{excurv}, the above becomes
\begin{eqnarray}
E^b{}^i
\HH_{a b\,i 0}
&=& 
\pi_\phi \partial_{a}\phi
+
2\,E^b{}^i
\ovr{A}_{a\, i 0}\,\phi\,\partial_{b}\phi
-
\frac{1}{\beta}\,
E^b{}^i
\epsilon_{ijk}
K_{a j} K_{b k}
\nppp
&&
+
E^b{}^i
\partial_{a} (\phi^2 \ovr{A}_{b\, i 0})
+
E^b{}^i
\ovr{A}_{a\, i j} (\phi^2  \ovr{A}_{b j 0})
-
E^b{}^i
\partial_{b} (\phi^2 \ovr{A}_{a\, i 0})
-
E^b{}^i
\ovr{A}_{b\, i j} (\phi^2  \ovr{A}_{a j 0}).
\label{fff2a}
\end{eqnarray}
By using
\begin{eqnarray}
\ovr{A}_{a i 0}
&=&
\frac{1}{\beta}\,\phi^{-2}
(A_{a i} - \Gamma_{a i})
\nppp
&=&
\phi^{-2}C_{a i}
\label{iAai}
\end{eqnarray}
%
from \eqref{Gai}, \eqref{Aai}, and \eqref{AAai},
and
\begin{eqnarray}
\ovr{A}_{a i j}
&=&
\Gamma_{a i j}
+
\phi_{a i j}
\nppp
&=&
\Gamma_{a i j}
+
\phi^{-1}
\phi_{,c}\, (
e^c{}_j \,e_a{}_i
-
e^c{}_i \, e_a{}_j )
\label{iAaij}
\end{eqnarray}
from \eqref{bwab} and \eqref{ppaij}, Eq. \eqref{fff2a} becomes
\begin{eqnarray}
E^b{}^i
\HH_{a b\,i 0}
&=&
\pi_\phi \phi_{,a}
+
2\,E^b{}^i
C_{a i}\,\phi^{-1}\phi_{,b}
+
E^b{}^i
(
\phi_{a i j} C_{b j}
-
\phi_{b i j} C_{a j}
)
\nppp
&&
+
E^b{}^i
(
\nabla_a C_{b i}
-
\nabla_b C_{a i}
)
-
\frac{1}{\beta}\,
E^b{}^i
\epsilon_{ijk}
K_{a j} K_{b k}.
\label{EFabi0}
\end{eqnarray}
%



From \eqref{Kiax} we have
\begin{eqnarray}
K_{a i}
&=&
\phi^{-2}C_{a i}
-
\frac{\sqrt{E}}{N}\,\phi^{-1}(\dot\phi-N^c\phi_{,c})\,E_{a i}
\nppp
&=&
\phi^{-2}C_{a i}
-
\phi_{a i 0}
\label{EKia}
\end{eqnarray}
yielding
\begin{eqnarray}
\epsilon_{ijk}
E^b{}^i K_{a j}K_{b k}
&=&
\phi^{-4}
C_{a j}
\epsilon_{ijk} C_{b k} E^b{}^i
-
\frac{\sqrt{E}}{N}\,\phi^{-3}(\dot\phi-N^c\phi_{,c})\,
E_{a j}
\epsilon_{ijk} C_{b k} E^b{}^i
\nppp
&=&
\phi^{-4}
C_{a j}\,\GG_j
-
\frac{\sqrt{E}}{N}\,\phi^{-3}(\dot\phi-N^c\phi_{,c})\,
E_{a j}\,\GG_j.
\label{EKK}
\end{eqnarray}

Therefore by using \eqref{EKK}, Eq. \eqref{EFabi0} becomes
\begin{eqnarray}
E^b{}^i
\HH_{a b\,i 0}
&=& 
E^b{}^i
(
\nabla_a C_{b i}
-
\nabla_b C_{a i}
)
+
\pi_\phi \phi_{,a}
\nppp
&&
+
2\,\phi^{-1}\phi_{,c}\,E^c{}^i
C_{a i}
+
E^b{}^i
(
\phi_{a i j} C_{b j}
-
\phi_{b i j} C_{a j}
)
\label{EFabi0a}
\end{eqnarray}
up to adding terms proportional to the Gauss constraint.

From \eqref{ppaij} we have
\begin{eqnarray*}
\phi_{a i j} E^b{}^i C_{b j}
&=&
\phi^{-1}
\phi_{,c}\, (
E^c{}_i \, C_{a i}
-
E_a{}_i \, C^c{}_i )
\ppp
\phi_{b i j} E^b{}^i C_{a j}
&=&
2\,\phi^{-1}
\phi_{,c}\,
E^c{}^i C_{a i}.
\end{eqnarray*}
%


Substituting the above relations into \eqref{EFabi0a}, we obtain
\begin{eqnarray}
E^b{}^i
\HH_{a b\,i 0}
&=& 
2\,
\nabla_{[a}C_{b]\, i}\,E^b{}^i
+
\phi_{,a} \pi_\phi
\label{EFabi0b}
\end{eqnarray}
up to adding terms proportional to the Gauss constraint.

Comparing  \eqref{EFkab2} and \eqref{EFabi0b}, we see that subject to the Gauss constraint, the weakly vanishing of \eqref{EFabi0b} is equivalent to the weakly vanishing of
\begin{eqnarray}
\HH_a
&=&
\beta^{-1}F_{ab\,i} E^{b i}
+
\phi_{,a} \pi_\phi
\label{diffeomcons}
\end{eqnarray}
which is identified as the {\it diffeomorphism constraint}.

\subsubsection{Hamiltonian constraint}

Using \eqref{A2Gam}, along with \eqref{FP},
\eqref{Gai}, \eqref{excurv}, \eqref{iAai}, and \eqref{iAaij},
the third term in Eq. \eqref{fff} gives
\begin{eqnarray*}
\phi^{-2} \HH_{a b\,i j}
&=& 
R_{a b\,i j}
+
\partial_{a} \phi_{b\, i j}
+
\Gamma_{a\, i k} \phi_{b\, k j}
+
\Gamma_{a\, j k} \phi_{b\, i k}
-
\partial_{b} \phi_{a\, i j}
-
\Gamma_{b\, i k} \phi_{a\, k j}
-
\Gamma_{b\, j k} \phi_{a\, i k}
\nppp
&&
+
\phi_{a\, i k} \phi_{b\, k j}
-
\phi_{b\, i k} \phi_{a\, k j}
+
\phi^{-4}
(
C_{a i} C_{b j}
-
C_{b i} C_{a j}
)
\nppp
&&
+
\beta^{-1}
\epsilon_{ijk}
\phi^{-2}
(
\partial_{a} K_{b k}
-
\partial_{b} K_{a k}
+
\Gamma_{a\, k m} K_{b m}
-
\Gamma_{b\, k m} K_{a m}
).
\end{eqnarray*}

Using \eqref{EKia}, we have
\begin{eqnarray*}
\partial_{a} K_{b k}
+
\Gamma_{a\, k m} K_{b m}
&=&
\phi^{-2}
(
\partial_{a} C_{b k}
+
\Gamma_{a\, k m} C_{b m}
)
\nppp
&&
-
(
\partial_{a} \phi_{b k 0}
+
\Gamma_{a\, k m} \phi_{b m 0}
)
-
2\phi^{-3}\phi_{,a} C_{b k}.
\end{eqnarray*}

Substituting this into the above, we have
\begin{eqnarray*}
\phi^{-2} E^a{}^i  E^b{}^j \HH_{a b\,i j}
&=& 
E^a{}^i  E^b{}^j
R_{a b\,i j}
+
E^a{}^i  E^b{}^j
\big(
2\nabla_{[a} \phi_{b]\, i j}
+
\phi_{a\, i k} \phi_{b\, k j}
-
\phi_{b\, i k} \phi_{a\, k j}
\big)
+
2\phi^{-4}
E^a{}^i  E^b{}^j
C_{[a}{}^{i} C_{b]}{}^{j}
\nppp
&&
+
2\,\beta^{-1}E^a{}^i  E^b{}^j
\epsilon_{i j k}
\big(
\phi^{-4} \nabla_{[a} C_{b] k}
+
\phi^{-5} \phi_{,b} C_{a k}
-
\phi^{-5} \phi_{,a} C_{b k}
\big)
\nppp
&&
-
2\,\beta^{-1}\phi^{-2}
\epsilon_{i j k}
\nabla_{[a} (\phi_{b] k 0} E^a{}^i  E^b{}^j).
\end{eqnarray*}

From \eqref{ppai0} we have
\begin{eqnarray*}
2\epsilon_{i j k}
\nabla_{[a} (\phi_{b] k 0} E^a{}^i  E^b{}^j)
&=&
\epsilon_{i j k}
\nabla_{a}
\big[\frac{\sqrt{E}}{N\,\phi}\,(\dot\phi-N^c\phi_{,c})\,E^a{}^i \delta^j_k\big]
-
\epsilon_{i j k}
\nabla_{b}
\big[\frac{\sqrt{E}}{N\,\phi}\,(\dot\phi-N^c\phi_{,c})\,E^b{}^j \delta^i_k\big]
=
0.
\end{eqnarray*}

Applying the above relation, we have
\begin{eqnarray}
E^a{}^i  E^b{}^j \HH_{a b\,i j}
&=& 
-\phi^{2} \epsilon_{ijk}
E^a{}^i  E^b{}^j
R_{ab\,k}
+
2\phi^{-2}
E^a{}^i  E^b{}^j
C_{[a}{}^{i} C_{b]}{}^{j}
\nppp
&&
+
\phi^{2} E^a{}^i  E^b{}^j
\big(
2\nabla_{[a} \phi_{b]\, i j}
+
\phi_{a\, i k} \phi_{b\, k j}
-
\phi_{b\, i k} \phi_{a\, k j}
\big)
\nppp
&&
-
2\,\beta^{-1}\phi^{-2}
E^a{}^i
\nabla_{a}\,\GG_i
\label{fff3a}
\end{eqnarray}

Using the identity \eqref{epsiFPP1x} and relations
\begin{gather*}
E^a{}^i  E^b{}^j
\nabla_{[a}\phi_{b] i j}
=
2E\phi^{-2}\phi_{,a}\phi^{,a}
-
2E\phi^{-1}\Delta\phi
\ppp
E^a{}^i  E^b{}^j
\phi_{[a}{}^{i k} \phi_{b]\, k j}
=
-E\phi^{-2}\phi_{,a}\phi^{,a}
\end{gather*}
where $\Delta$ is the Laplace-Beltrami operator of $h_{ab}$,
obtained from \eqref{ppaij}, and neglecting terms proportional to $\nabla_{a}\,\GG_i$, Eq. \eqref{fff3a} then yield the {\it Hamiltonian constraint} from the third term of \eqref{Lha} divided by $(-N)$ as follows.
\begin{eqnarray}
\HH
&=& 
\frac{\phi^2}{2\sqrt{E}}
\epsilon_{i j k}
F_{ab\,k}
E^a{}^i E^b{}^j
-
\frac{1}{\sqrt{E}\phi^2}
\big(
\beta^2\phi^4 + 1
\big)
C_{[a}{}^{i} C_{b]}{}^{j}
E^a{}^i E^b{}^j
\nppp
&&
+
\sqrt{E}\big(
2\phi\Delta\phi
-
\phi_{,a}\phi^{,a}
\big).
\label{hamcons}
\end{eqnarray}
%

\subsection{Conformal transformation properties}

Consider the simultaneous conformal transformations
\begin{eqnarray}
\phi
&\rightarrow&
\Omega^{-1}\phi
\label{omga}
\ppp
E^a{}_i
&\rightarrow&
\Omega^2 E^a{}_i
\label{omgb}
\end{eqnarray}
for any positive function $\Omega(x)$.

From \eqref{ppai0} and \eqref{excurv}, we have
\begin{subequations}
\begin{eqnarray}
K_{ai}
&\to&
K_{ai}
+
\frac{\sqrt{E}}{N}\,
\Omega^{-1}
(\dot\Omega-N^c\Omega_{,c})\,E_{a i}
\ppp
\phi_{a i 0}
&\to&
\phi_{a i 0}
-
\frac{\sqrt{E}}{N}\,
\Omega^{-1}
(\dot\Omega-N^c\Omega_{,c})\,E_{a i}.
\end{eqnarray}
\label{omgKphi}
\bppp
\end{subequations}

It follows from \eqref{piphi}, \eqref{Kiax} and \eqref{omgKphi} that
\begin{eqnarray}
C_{a i}
&\to&
\Omega^{-2}\,C_{a i}
\label{Kiax2}
\end{eqnarray}
and
\begin{eqnarray}
\pi_\phi
&\to&
\Omega\,\pi_\phi.
\label{omgpi}
\end{eqnarray}

From \eqref{Gai} we obtain
\begin{eqnarray}
{\Gamma}_a^i
&\to&
{\Gamma}_a^i
-
\frac{1}{4}\,
\Omega^{-1}
\Omega_{,b}\,
\epsilon_{ijk}
E_a{}_j E^b_k.
\label{Gai2}
\end{eqnarray}

Therefore, using \eqref{Kiax2} and \eqref{Gai2}, we find that
\begin{subequations}
\begin{eqnarray}
A_a^i
&\to&
A_a^i
+
\beta
(\Omega^{-2}-1)
C_a^i
-
\frac{1}{4}\epsilon_{ijk}
E_a^j E^b_k\,
(\ln\Omega)_{,b}
\ppp
&&\text{or equivalently}
\nppp
A_a^i
&\to&
A_a^i|_{\beta=1}
+
(\beta\Omega^{-2}-1)C_a^i
-
\frac{1}{4}\,
\epsilon_{ijk}
\Omega^{-1}
\Omega_{,b}\,
E_a^j E^b_k.
\end{eqnarray}
\label{omgc}
\bppp
\end{subequations}

\newpage
\section{Generalized ADM formalism with conformal Ashtekar-Barbero variables}
\label{sec:2}

\subsection{Canonical conformal analysis}



Consider the Lagrangian density for the Einstein gravity using the metric $\ovr{g}_{\mu\nu}$,
\begin{eqnarray}
\LLL
&=&
\frac1{2}\sqrt{-\ovr{g}}\,\ovr{R}
\label{L0}
\end{eqnarray}
where
$\ovr{g} = \det(\ovr{g}_{\mu\nu})$,
$\ovr{R}$ is the scalar curvature of $\ovr{g}_{\mu\nu}$,
and
\begin{eqnarray}
\ovr{g}_{\mu\nu}
&=&
\phi^2 g_{\mu\nu}
\label{cnf1}
\end{eqnarray}
%
where $\phi$ is a scalar field.

The following qualities are conformally transformed from the original (barred) quantities according to:
\begin{eqnarray}
\ovr{h}_{ab}
&=&
\phi^2 h_{ab}
\label{cnf2}
\\
\ovr{N}
&=&
\phi N
\label{cnf3}
\\
\ovr{N}^a
&=&
N^a.
\label{cnf4}
\end{eqnarray}

It follows that
\begin{eqnarray}
\ovr{h}^{ab}
&=&
\phi^{-2} h^{ab}
\label{cnf5x}
\\
\ovr{N}_a
&=&
\phi^2 N_a.
\label{cnf6}
\end{eqnarray}

Then, up to a total divergence Lagrangian density \eqref{L0} becomes
%
\begin{eqnarray}
\LLL
&=&
\frac{\phi^2}{2}\sqrt{h}N
\left(
K_{ab}K^{ab}-K^2+R
\right)
+
2\sqrt{h}\phi K
\left(
\dot{\phi} - N^a\phi_{,a}
\right)
-
2N \sqrt{h}(\phi\,\phi_{,a})^{;a}
\nppp&&
-
\frac{3}{N}\sqrt{h}
\left[
\dot{\phi}^2
-2N^a\dot{\phi}\phi_{,a}
-\left(N^2h^{ab}-N^aN^b\right)\phi_{,a}\phi_{,b}
\right].
\label{eqL}
\end{eqnarray}

The canonical momenta of the metric and scalar fields follow respectively from \eqref{eqL} to be
\begin{eqnarray}
p^{ab}
&=&
\frac{\partial\LLL}{\partial \dot{h}_{ab}}
=
-
\frac1{N}\sqrt{h}\,h^{ab}\phi\,(\dot{\phi}-N^c\phi_{,c})
-
\frac{\phi^2}{2}\sqrt{h}\,(K^{ab}-h^{ab}K)
\label{ppab0}
\ppp
\pi_\phi
&=&
\frac{\partial\LLL}{\partial\dot{\phi}}
=
-
\frac{6}{N}\sqrt{h}\,(\dot{\phi}-N^c\phi_{,c})
+
2 \sqrt{h}\,\phi K.
\label{piphi0}
\end{eqnarray}
%


From \eqref{defKij} we have
\begin{eqnarray}
\dot{h}_{ab}
=
-2N K_{ab} + N_{a;b} + N_{b;a}.
\label{hij0}
\end{eqnarray}

From \eqref{ppab0} we have
\begin{eqnarray}
p
&=&
-
\frac{3}{N}\sqrt{h}\,\phi\,(\dot{\phi}-N^c\phi_{,c})
+
\phi^2\sqrt{h}\,K.
\label{eq43p}
\end{eqnarray}

From \eqref{piphi0} we have
\begin{eqnarray*}
- \frac{6}{N}\sqrt{h}\,(\dot{\phi}-N^c\phi_{,c})
=
\pi_\phi - 2 \sqrt{h}\,\phi K
\end{eqnarray*}
and therefore
\begin{eqnarray}
\dot{\phi}
&=&
N^c\phi_{,c}
-
\frac{N}{6\sqrt{h}}\pi_\phi + \frac{N}{3} \phi K.
\label{eqphid}
\end{eqnarray}

Substituting \eqref{eqphid} into \eqref{ppab0} we get
\begin{eqnarray}
p^{ab}
&=&
-
\frac{1}{6}\,h^{ab}\phi
\left(
\pi_\phi - 2 \sqrt{h}\,\phi K
\right)
-
\frac{\phi^2}{2}\sqrt{h}(K^{ab}-h^{ab}K)
\nppp
&=&
\frac{1}{6}\,h^{ab}\phi
\pi_\phi
-
\frac{\phi^2}{2}\sqrt{h}
\left(K^{ab}-\frac13\,h^{ab}K\right).
\label{pab}
\end{eqnarray}

Hence we have
\begin{eqnarray}
K^{ab}-\frac13 h^{ab}K
&=&
-
\frac{2}{\phi^2\sqrt{h}}
\left(
p^{ab}
-\frac16 h^{ab}\phi
\pi_\phi
\right).
\label{eqKij0}
\end{eqnarray}

By contracting with $h_{ab}$ the above becomes
\begin{eqnarray}
\phi\,\pi_\phi - 2p
=0
\label{cf0}
\end{eqnarray}
where
\begin{eqnarray*}
p &=& h_{ab}p^{ab}.
\end{eqnarray*}

For later quantization, the above equation will be treated as a weak condition $\CC\approx0$ by introducing the conformal constraint
\begin{eqnarray}
\CC
&=&
\phi\,\pi_\phi - 2p.
\label{cf}
\end{eqnarray}

Using \eqref{cf0}, we can rewrite \eqref{eqKij0} as
\begin{eqnarray}
K^{ab}-\frac13 h^{ab}K
&=&
-
\frac{2}{\phi^2\sqrt{h}}
\left(
p^{ab}
-\frac13 h^{ab} p
\right).
\label{eqKij0a}
\end{eqnarray}

Using \eqref{eqKij0a}, we can then rewrite \eqref{hij0} as
\begin{eqnarray}
\dot{h}_{ab}
&=&
\frac{4 N}{\phi^2\sqrt{h}}
\left(
p_{ab}
-\frac13 h_{ab} p
\right)
-\frac{2}{3} N h_{ab} K
+ N_{a;b} + N_{b;a}.
\label{hij02}
\end{eqnarray}

It follows that
\begin{eqnarray}
K_{ab}K^{ab}
&=&
\frac{4}{\phi^4{h}}
\left(
p_{ab}
p^{ab}
-
\frac13 p^2
\right)
+
\frac13\, K^2
\label{eqKK0}
\end{eqnarray}

Substituting \eqref{eqphid}, \eqref{cf0}, \eqref{hij02} and \eqref{eqKK0} into \eqref{eqL}, up to a total divergence, we have
\begin{eqnarray*}
\HHH
&=&
p^{ab}\dot{h}_{ab}+\pi_\phi\dot{\phi}-\LLL
\npppp
&=& 
-2N^a p^{b}{}_{a;b}
+N^a\phi_{,a}\pi_\phi
+
\frac{2 N}{\phi^2\sqrt{h}}
\left(
p_{ab}p^{ab}
-\frac13 p^2
\right)
-
\frac{\phi^2}{2}\sqrt{h}\,N\,R
\nppp&&
-\frac{N \pi_\phi}{24\phi\sqrt{h}}
-
3N \sqrt{h}\,h^{ab}\phi_{,a}\phi_{,b}
+
2N \sqrt{h}\,(\phi\,\phi_{,a})^{;a}
\end{eqnarray*}

Furthermore, the above becomes
\begin{eqnarray*}
\HHH
&=&
N \HH + N^a\HH_a
\end{eqnarray*}
where we have introduced the Hamiltonian constraint
\begin{eqnarray}
\HH
&=&
\frac{2}{\phi^2\sqrt{h}}\,
\big(
p_{ab}p^{ab} - \frac12\, p^2
\big)
-
\frac{\phi^2}{2}\sqrt{h}\,R
+
\sqrt{h}\,
\big(
2\,\phi\Delta \phi
-
\phi_{,a}\phi^{,a}
\big)
\label{Chamil}
\end{eqnarray}
and the diffeomorphism constraint
\begin{eqnarray}
\HH_a
&=&
-2p^{b}{}_{a;b}+\phi_{,a}\pi_\phi
\label{Cdiff}
\end{eqnarray}
supplemented with the conformal constraint \eqref{cf}.

\subsection{Canonical transformation to the triad variables}

We use $i,j,\cdots=1,2,3$ as triad indices, so that
\begin{eqnarray}
h_{ab}
&=&
\delta_{ij} e^i_a e^j_b
=
h\delta_{ij} E^i_a E^j_b
\label{hab}
\end{eqnarray}
where
$E^a_i = \sqrt{h}\, e^a_i$
is the densitized triad with $E^i_a$ as the inverse.

Using \eqref{hab} we have
\begin{eqnarray}
\dot{h}_{ab}
&=&
\frac1h (h_{ab} h_{cd} -h_{ac}h_{bd}-h_{ad}h_{bc})E^c_i \dot{E}^d_i.
\label{hab0a}
\end{eqnarray}

It follows from \eqref{pab}, \eqref{eqKij0}  and \eqref{hab0a} that
\begin{eqnarray}
p^{ab}\dot{h}_{ab}
&=&
-
\left[
-
\frac{1}{6}\,h^{ab}\phi
\pi_\phi
+
\frac{\phi^2}{2}\sqrt{h}
\left(K^{ab}-\frac13\,h^{ab}K\right)
\right]
\frac1h (h_{ab} h_{cd} -h_{ac}h_{bd}-h_{ad}h_{bc})E^c_i \dot{E}^d_i
\nppp
&=&
- C^i_a \dot{E}^a_i
=
E^a_i \dot{C}^i_{a}
- (E^a_i C^i_a)_{,t}
\label{KEcano}
\end{eqnarray}
where
\begin{eqnarray}
C^i_a
&=&
\frac{1}{h} (2 p_{ab} - h_{ab} p) E^b_i.
\label{Kia}
\end{eqnarray}

Substituting \eqref{ppab0} and \eqref{eq43p} into \eqref{Kia}, we find
\begin{eqnarray}
C^i_a
&=&
\phi^2 K^i_a
+
\phi\,(\dot{\phi}-N^c\phi_{,c})
\frac{E_{a i}}{N \sqrt{h}}
\label{Kia2}
\end{eqnarray}
where
\begin{eqnarray}
K_{a i}
&=&
-K_{ab}e^b{}_i
\label{Ke2}
\end{eqnarray}
are the extrinsic curvature 1-forms as in \eqref{Ke}.

Contracting \eqref{Kia} with $E^a_i$ and using \eqref{hab}, we get
\begin{eqnarray}
C^i_a E^a_i
&=&
-p.
\label{pKE}
\end{eqnarray}
Substituting \eqref{pKE} into \eqref{Kia} and contracting with $E^i_d$, we have
\begin{eqnarray}
p_{ab}
&=&
\frac12\big(
h\, C^i_a E^i_b
-
h_{ab} C^k_c E^c_k
\big).
\label{pabKE0}
\end{eqnarray}

Since $p_{ab}=p_{ba}$, we see from \eqref{pabKE0} that $C^i_a E^i_b = C^i_b E^i_a$.
This implies a constraint $C^i_{[a} E^i_{b]}\approx0$ or equivalently
\begin{eqnarray}
\mathcal{C}_k = \epsilon_{kij}C_{a[i}E^a_{j]} = - \epsilon_{kij} C^l_{[a} E^l_{b]} E^a_i E^b_j
\label{spinconsx}
\end{eqnarray}

From \eqref{KEcano}, we see that the following variables then form canonical pairs:
\begin{eqnarray}
(C^i_a, E^a_i) \text{ and } (\phi, \pi_\phi).
\label{cano}
\end{eqnarray}

Using these canonical variables and \eqref{pKE}, we see that \eqref{cf} becomes
\begin{eqnarray}
\CC
&=&
\phi\,\pi_\phi + 2 C^i_{a} E^a_{i}.
\label{cf2}
\end{eqnarray}

\subsubsection{Expression for the Hamiltonian constraint}

Consider
\begin{eqnarray}
A^i_a
&=&
{\Gamma}^i_a
+
\beta C^i_a
\label{pia1}
\end{eqnarray}
with an Immirzi-type parameter $\beta$.

Since ${\Gamma}^i_a$ commutes with ${E}^a_i$,
we see that $(A^i_a, E^a_i)$ form canonical variables, complete with $(\phi,\pi_\phi)$.

From \eqref{epsiFPP1x}, see that
\begin{eqnarray}
R
&=&
-
h^{-1}
\big(
\epsilon_{i j k}
F^k_{ab}
E^a_i E^b_j
-
2\beta^2\,
C_{[a}^{i} C_{b]}^{j}
E^a_i E^b_j
\big)
\label{LCR}
\end{eqnarray}
up to adding a term proportional to ${\nabla}^{}_{a}\, \GG_i$.

From \eqref{pabKE0}, we have
\begin{eqnarray}
p_{ab}
p^{ab}
-\frac12\,p^2
&=&
-\frac{1}{2}
C^i_{[a} C^j_{b]} E^a_{i} E^b_{j}.
\label{ppKE}
\end{eqnarray}

By substituting \eqref{LCR} and \eqref{ppKE}  into \eqref{Chamil}, we obtain
\begin{eqnarray}
\HH
&=&
\frac{\phi^2}{2\sqrt{h}}\,
\epsilon_{kij} F^k_{ab} E_{i}^a E_{j}^b
-
\frac{1}{\phi^2\sqrt{h}}\,
\big(
1 + \phi^4\beta^2
\big)
C^i_{[a} C^j_{b]} E_i^a E_j^b
+
\sqrt{h}\,
\big(
2\,\phi\Delta \phi
-
\phi_{,a}\phi^{,a}
\big)
\label{Chamil3a}
\end{eqnarray}
up to adding terms proportional to the covariant derivative of the Gauss constraint.

\newpage
\section{Conformally augmented gauge theory of scale-invariant dilaton gravitation}
\label{sec:3}

Consider the total Lagrangian density
\begin{eqnarray}
\LLL
&=&
\LLL_\G[g_{\mu\nu},\phi]
+
\LLL_\S[g_{\mu\nu},\phi,\A_\mu]
+
\LLL_\M[g_{\mu\nu},\phi,\A_\mu,\psi]
\label{LLLtot}
\end{eqnarray}
with the following
\begin{eqnarray}
\LLL_\G
&=&
\frac12\,\Phi^2
\sqrt{-g}\,R
\label{LLLG}
\end{eqnarray}
where $R$ is the scalar curvature of $g_{\mu\nu}$ and
\begin{eqnarray}
\Phi^2
&=&
\sum_{A}\frac{\phi_A\phi^A}{6k_{A}}.
\label{Phi}
\end{eqnarray}
%
Some of the index $A$ are Haar measure indices associated with a gauge group $G$.
The sum over repeated $A$ would be often omitted like other indices when unambiguous.

The term
\begin{eqnarray}
\LLL_\S
&=&
-
\frac{1}{2}\sqrt{-g}\,
\DD_\mu\phi_A \DD^\mu\phi^A
-
\sqrt{-g}\, V(\phi)
\label{LLLS}
\end{eqnarray}
is the scalar Lagrangian, where $V(\phi)$ is a homogeneously fourth order potential in $\phi^A$, and $Dy_\mu$ is the covariant derivative using the LC connection and gauge connection $\A_\mu$.

The scalar fields $\phi_A$ have been normalized with a minus sign in the kinetic terms of \eqref{LLLS} for a stable positive kinetic energy. In principle, a negative sign is possible e.g. for a pure conformal $\phi$ as treated in preceding conformally treated GR theory. Parameter $k_A$ could in principle take any real value, with $k_A=1$ for any conformally coupled scalar $\phi_A$ and $|k_A| \to \infty$ for any minimally coupled scalar $\phi_A$.

The term
\begin{eqnarray}
\LLL_\M
&=&
\Re\,\sqrt{-g}\,
\tld{\psi}
\big[
\gamma^I e^\mu{}_I
\DD_\mu
+
\mu(\phi)
\big]
\psi
-
\frac14\,\sqrt{-g}\,
\FF_{\mu\nu}^A\FF^{\mu\nu}_A
\label{LLLM}
\end{eqnarray}
is the matter Lagrangian, using
$\FF_{\mu\nu}$ as the curvature of $\A_\mu$,
the constant Dirac matrices
$\gamma^I$
satisfying the Clifford algebra
$\gamma^{(I}\gamma^{J)} = \eta^{IJ}$
with the Hermiticity condition
$\gamma^I{}^\dag = \gamma^0\gamma^I\gamma^0$,
the adjoint spinsor
$\tld{\psi} = i\psi^\dagger\gamma^0$,
and a Yukawa coupling matrix
$\mu(\phi)$
homogeneous linearly in $\phi^A$ \cite{Bekenstein1980}.

Lagrangian \eqref{LLLtot} has the property of being invariant under the following constant conformal transformations:
\begin{subequations}
\begin{eqnarray}
\phi
&\rightarrow&
\Omega^{-1}\phi
\label{Omga}
\ppp
g_{\mu\nu}
&\rightarrow&
\Omega^2 g_{\mu\nu}
\label{Omgb}
\ppp
\A_\mu
&\rightarrow&
\A_\mu
\label{Omgc}
\ppp
\psi
&\rightarrow&
\Omega^{-3/2}\psi
\label{Omgd}
\end{eqnarray}
\label{scltrans}
\bppp
\end{subequations}
for any positive constant $\Omega$.

\subsection{Conserved Weyl current}

The scaling symmetry implies the existence of a conserved Noether current, which can be identified from the boundary terms from the on shell variation of the Lagrangian density \eqref{LLLtot} under an infinitesimal scaling transformation with $\Omega=1+\varepsilon$ in \eqref{scltrans} for an infinitesimal $\varepsilon$, denoted by
$\delta_\varepsilon$.

To this end, first it can be obtained from \eqref{LLLG} that
\begin{eqnarray*}
\delta_\varepsilon
\LLL_\G
&=&
\frac12\,
\sqrt{-g}\,
{\nabla}_\mu
\big(
\Phi^2
g^{\alpha\beta}\delta_\varepsilon\Gamma^\mu_{\alpha\beta}
-
\Phi^2
g^{\mu\beta}\delta_\varepsilon\Gamma^\alpha_{\alpha\beta}
\big)
\nppp
&=&
0
\end{eqnarray*}
where $\nabla_\mu$ is the LC covariant derivative using $g_{\mu\nu}$,
on account of the invariance of the LC connection under scaling transformations, and hence $\delta_\varepsilon\Gamma^\mu_{\alpha\beta}=0$.

A nontrivial contribution is derived from \eqref{LLLS} as follows
\begin{eqnarray}
\delta_\varepsilon
\LLL_\S
&=&
-
\sqrt{-g}\, {\nabla}_\mu
\Big(
g^{\mu\nu}
\partial_\nu\phi_A
\delta_\varepsilon\phi^A
\Big)
\nppp
&=&
\frac{\epsilon}{2}\,
\sqrt{-g}\, {\nabla}_\mu
\partial^\mu\phi^2
\label{seqc}
\end{eqnarray}
where
\begin{eqnarray*}
\phi^2 = \phi_A\phi^A
\end{eqnarray*}
since $\delta_\varepsilon \phi=-\varepsilon \phi$.

There are no additional relevant boundary terms as a result of varying gauge connections since they are invariant under the scale transformations, i.e. $\delta_\varepsilon \A_\mu=0$.

Furthermore, by varying $\eqref{LLLM}$, one finds that the scale transformations of spinors under \eqref{Omgd} do not contribute to any boundary term either, as
$\delta_\varepsilon\LLL_\G=0$.

It therefore follows from the above discussions, especially \eqref{seqc}, that the scale invariance relation $\delta_\epsilon\LLL=0$ gives to the conserved {\it Weyl current}
$\partial_\mu\phi^2$.

\subsection{Canonical analysis}

To focus on the gravitational structure with multiple scalar fields, in the following we suppress the gauge field $\A_\mu$, spinor $\psi$, and the scalar potential $V(\phi)$, so that
\begin{eqnarray}
\DD_\mu\phi^A
&=&
\partial_\mu\phi^A
\end{eqnarray}
and
\begin{eqnarray}
\LLL
&=&
\LLL_\G+\LLL_\S
\label{LLLtot2}
\end{eqnarray}
with
\begin{eqnarray}
\LLL_\S
&=&
-
\frac{1}{2}\sqrt{-g}\,
\partial_\mu\phi_A \partial^\mu\phi^A.
\label{LLLS2}
\end{eqnarray}

The Lagrangian density $\LLL$ up to a total divergence becomes
\begin{eqnarray}
\LLL
&=&
\frac{\Phi}{2}\sqrt{h}N
\big(K_{ab}K^{ab}-K^2+R\big)
-
N\sqrt{h}\,{\Delta}\Phi
-
\sqrt{h}\,K N^a\Phi_{,a}
+
\sqrt{h}\,K \dot{\Phi}
\nppp&&
-
\frac{\sqrt{h}}{2 N}
\Big[
\dot{\phi}_A\dot{\phi}^A
-
2 N^a (\partial_a\phi_A)\dot{\phi}^A
-
\big(N^2h^{ab}-N^a N^b\big)
\partial_a\phi_A\,\partial_b\phi^A
\Big].
\label{eqLx}
\end{eqnarray}

As a generalization of \eqref{ppab0} and \eqref{piphi0}, here the canonical momenta of the metric and scalar fields follow respectively from \eqref{eqLx} to be
\begin{eqnarray}
p^{ab}
&=&
-
\frac{\sqrt{h}}{N}\,h^{ab}\Phi(\dot{\Phi}-N^c\Phi_{,c})
-
\frac{\Phi^2\sqrt{h}}{2}\,(K^{ab}-h^{ab}K)
\label{ppab}
\ppp
\pi_A
&=&
-
\frac{\sqrt{h}}{N}
\big[
\dot{\phi}_A
-
N^a \partial_a\phi_A
\big]
+
\frac{\sqrt{h}}{3k_A}\, K \phi_A.
\label{piphi1}
\end{eqnarray}
%

%

Substituting \eqref{eqphid} into \eqref{ppab} we get
\begin{eqnarray}
K^{ab}
&=&
\frac{h^{ab}K}{3}
-
\frac{2}{\Phi^2\sqrt{h}}
\Big(
p^{ab}
-
\frac{h^{ab} p}{3}
\Big)
\label{KKab2}
\end{eqnarray}

From \eqref{ppab} we have
\begin{eqnarray}
p
&=&
-
\frac{3\sqrt{h}}{N}\,\Phi(\dot{\Phi}-N^c\Phi_{,c})
+
\Phi^2\sqrt{h}\,K
\label{ppabp}
\end{eqnarray}
yielding
\begin{eqnarray}
\dot{\Phi}-N^c\Phi_{,c}
&=&
\frac{N K}{3}
-
\frac{N p}{3\Phi\sqrt{h}}.
\label{eqphidx}
\end{eqnarray}

Let us introduce the conformal constraint
\begin{eqnarray}
\CC
&:=&
\phi^A \pi_{A} - 2 p.
\label{cfc}
\end{eqnarray}

From \eqref{piphi1} and \eqref{KKab2} we find the relation
\begin{eqnarray}
K \sqrt{h}\sum_{A}
\frac{k_A-1}{6k_A^2}\,\phi_A\phi^A
&=&
\sum_{A}
\frac{k_A-1}{2k_A}\,\pi_A \phi^A
-
\frac{\CC}{2}.
\label{Kppp4}
\end{eqnarray}

Substituting \eqref{piphi1}, \eqref{defKij}, \eqref{KKab2}, and  \eqref{eqphidx} into \eqref{eqLx}, up to a total divergence, we obtain
\begin{eqnarray}
\HHH
&=&
N \HH + N^a\HH_a
\end{eqnarray}
using the diffeomorphism constraint
\begin{eqnarray}
\HH_a
&=&
\partial_a\phi^A\, \pi_A
-
2\,{\nabla}_b\, p^{b}{}_{a}
\label{Cdiffx}
\end{eqnarray}
where $\nabla_a$ is the LC covariant derivative using $h_{ab}$,
and the Hamiltonian constraint
\begin{eqnarray}
\HH
&=&
\frac{2}{\Phi^2\sqrt{h}}
\Big(
p_{ab}p^{ab}
-
\frac{p^2}{2}
\Big)
-
\frac{\Phi^2}{2}\sqrt{h}\,{R}
+
\sqrt{h}\,\Delta\Phi^2
\nppp&& 
-
\frac{\sqrt{h}}{2}
\partial_a\phi_A \partial^a\phi^A
+
\frac{p^2}{3\Phi^2\sqrt{h}}
-
\frac{\pi_A\pi^A}{2\sqrt{h}}
-
\sqrt{h}\, K^2
\sum_A
\frac{k_A-1}
{18k_A^2}\,\phi_A\phi^A.
\label{HHHa}
\end{eqnarray}

If all the scalar fields are conformally coupled, then
$
k_A
=
1
$
for all $A$.

In this case, Eq. \eqref{cfc} with $\CC=0$ allows \eqref{HHHa} to become
\begin{eqnarray}
\HH
&=&
\frac{2}{\Phi^2\sqrt{h}}
\Big(
p_{ab}p^{ab}
-
\frac{p^2}{2}
\Big)
-
\frac{ \Phi^2}{2}\sqrt{h}\,{R}
+
\sqrt{h}\,{\Delta}\Phi^2
\nppp&&
-
\frac{\sqrt{h}}{2}
\partial_a\phi_A \partial^a\phi^A
+
\frac{1}{12\Phi^2\sqrt{h}}\,
\phi^A \phi^B \pi_A \pi_B
-
\frac{\pi_A\pi^A}{2\sqrt{h}}.
\label{HHHx2}
\end{eqnarray}

For a single scalar field, Eq. \eqref{HHHx2} reduces simply to
\begin{eqnarray}
\HH
&=&
\frac{2}{\Phi^2\sqrt{h}}
\Big(
p_{ab}p^{ab}
-
\frac{p^2}{2}
\Big)
-
\frac{\Phi^2}{2}\sqrt{h}\,{R}
+
\sqrt{h}\,{\Delta}\Phi^2
-
\frac{\sqrt{h}}{2}
\partial_a\phi \partial^a \phi
\label{HHHxc}
\end{eqnarray}
where the scalar field becomes non-dynamical.

On the other hand, if at least one scalar field is non-conformally coupled, then by substituting \eqref{Kppp4} into \eqref{HHHa}, we get
\begin{eqnarray}
\HH
&=&
\frac{2}{\Phi^2\sqrt{h}}
\Big(
p_{ab}p^{ab}
-
\frac{p^2}{2}
\Big)
-
\frac{\Phi^2}{2}\sqrt{h}\,{R}
+
\sqrt{h}\,{\Delta}\Phi^2
-
\frac{\sqrt{h}}{2}
\partial_a\phi_A \partial^a \phi^A
\nppp&&
+
\frac{p^2}{3\Phi^2\sqrt{h}}
-
\frac{\pi_A\pi^A}{2\sqrt{h}}
-
\frac{2}{\sqrt{h}}
\Big(
p
-
\sum_A
\frac{\pi_A \phi_A}{2k_A}
\Big)^2
\Big(\sum_A
\frac{k_A-1}{k_A^2}\,\phi_A^2
\Big)^{-1}.
\label{hhha}
\end{eqnarray}

In the case of one set of non-conformal scalar fields $\phi_A$ with a common $k_A=k$ for all $A$, Eq. \eqref{hhha} reduces to
\begin{eqnarray}
\HH
&=&
\frac{2}{\Phi^2\sqrt{h}}
\Big(
p_{ab}p^{ab}
-
\frac{p^2}{2}
\Big)
-
\frac{\Phi^2}{2}\sqrt{h}\,{R}
+
\sqrt{h}\,{\Delta}\Phi^2
\nppp&& 
-
\frac{\sqrt{h}}{2}
\partial_a\phi_A \partial^a\phi^A
-
\frac{\CC^2}{12(k-1)\Phi^2\sqrt{h}}
\nppp&& 
+
\frac{1}{12k(k-1)\Phi^2\sqrt{h}}
\Big(
\phi^A \pi^B \phi_A \pi_B
-
\phi^A \phi^B \pi_A \pi_B
\Big).
\label{hhhc}
\end{eqnarray}

For a single scalar field, the last term of \eqref{hhhc} vanishes and \eqref{hhhc} recovers $\HH$ in \cite{Veraguth2017}.

Finally, by using \eqref{LCR} and \eqref{ppKE}, the above expressions are readily transcribed with the connection variable \eqref{pia1}, with other constraints taking the similar corresponding forms.

\subsection{Emergence of time}

The conserved Weyl current $\partial_\mu\phi^2$ suggests $\phi^2$  may be a natural candidate as a ``pre-'' or ``fuzzy'' harmonic time so that the physical wave functional
\begin{eqnarray}
\Psi[\psi, \A_a, A_{a}{}^{i}, \hat{\phi}\, ; \phi^2]
\label{PsiPhi}
\end{eqnarray}
in terms of the normalized scalar fields
$$
\hat{\phi}^A
=
\frac{\phi^A}{\sqrt{\phi^2}}.
$$

Although here we have separated out $\phi^2$, the function $\phi^2$ need not be used as a deparametrized functional time. Instead, we could start with a timeless Wheeler-DeWitt type equation $\HH \Psi=0$ for \eqref{PsiPhi} and only recover $\phi^2$ as cosmological time in the semiclassical limit.